\documentclass[sigconf]{acmart}

\AtBeginDocument{%
  \providecommand\BibTeX{{%
    \normalfont B\kern-0.5em{\scshape i\kern-0.25em b}\kern-0.8em\TeX}}}

\usepackage[pagewise]{lineno}
\usepackage{times}
\usepackage{fancyhdr,graphicx}
\usepackage{algorithm}
\usepackage{tikz}
\usepackage{algpseudocode}
\usepackage{listings}
\usepackage{tikz} 
\usepackage{hyperref}       
\usepackage{url}

\usepackage{cleveref}
\usepackage{caption}
\usepackage{subcaption}
\setcopyright{acmcopyright}
\copyrightyear{2023}
\acmYear{2023}
\acmDOI{XXXXXXX.XXXXXXX}

\acmConference[XXXX]{X}{2023}{X}

\newcommand{\CEL}{\textbf{CEL}}
\newcommand{\SWS}{\textbf{SWS}}
\newcommand{\PC}{\textbf{PC}}
\newcommand{\CA}{\textbf{CA}}
\newcommand{\RL}{\textbf{RL}}

\begin{document}

\title[ANTM: An Aligned Neural Topic Model For Exploring Evolving Topics]{ANTM: An Aligned Neural Topic Model \\For Exploring Evolving Topics}

\titlenote{\href{https://github.com/hamedR96/ANTM}{https://github.com/hamedR96/ANTM}}
\author{Hamed Rahimi}
 \authornote{hamed.rahim@sorbonne-universite.fr} 
 \affiliation{
   \institution{Sorbonne University}
   \city{Paris}
   \country{France}
}

\author{Hubert Naacke}
 \affiliation{
   \institution{Sorbonne University}
   \city{Paris}
   \country{France}
}

\author{Camelia Constantin}
 \affiliation{
   \institution{Sorbonne University}
   \city{Paris}
   \country{France}
}

\author{Bernd Amann}
 \affiliation{
   \institution{Sorbonne University}
   \city{Paris}
   \country{France}
}

\begin{abstract}
This paper presents an algorithmic family of dynamic topic models called Aligned Neural Topic Models (ANTM), which combine novel data mining algorithms to provide a modular framework for discovering evolving topics. ANTM maintains the temporal continuity of evolving topics by extracting time-aware features from documents using advanced pre-trained Large Language Models (LLMs) and employing an overlapping sliding window algorithm for sequential document clustering. This overlapping sliding window algorithm identifies a different number of topics within each time frame and aligns semantically similar document clusters across time periods. This process captures emerging and fading trends across different periods and allows for a more interpretable representation of evolving topics. Experiments on four distinct datasets show that ANTM outperforms probabilistic dynamic topic models in terms of topic coherence and diversity metrics. Moreover, it improves the scalability and flexibility of dynamic topic models by being accessible and adaptable to different types of algorithms. Additionally, a Python package is developed for researchers and scientists who wish to study the trends and evolving patterns of topics in large-scale textual data.

\end{abstract}



\begin{teaserfigure}
\centering
\includegraphics[width=0.9\textwidth]{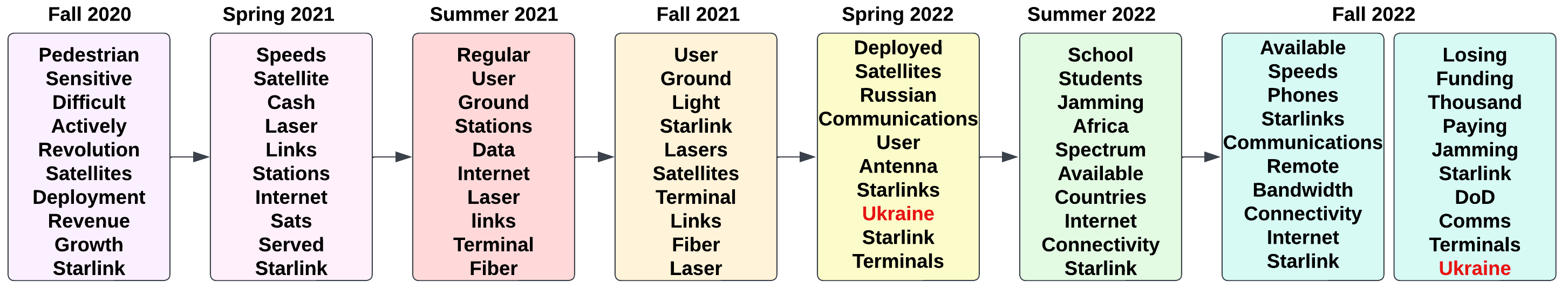}
\caption{Evolution of Elon Musk's Tweets about Starlink. \textnormal{The words Russia and Ukraine are appeared in spring and fall 2022 that coincide with two attacks of Russia on Ukraine and the tweets of Elon Musk in support of Ukrainians with the free Internet through Starlink.}}
\label{fig:musk}
\end{teaserfigure}

\received{2 Feb 2023}
\received[revised]{2 June 2023}
\maketitle

\section{Introduction}
Topic modeling is a statistical technique used in natural language processing to discover abstract themes from a corpus of text documents\cite{blei2012probabilistic, alghamdi2015survey}. These models are widely used in exploratory data analysis for organizing, understanding, and summarizing large amounts of text data~\cite{churchill2022evolution}. Dynamic Topic Models~\cite{blei2006dynamic,ABDELRAZEK2023102131} are a variant of topic models for temporal archives. These models update their estimates of the underlying topics as new documents are added to the corpus, allowing the identification of evolving trends and patterns in the archive. Topic evolution analysis~\cite{wang2006topics,hu2015modeling} has been used in a variety of applications, such as discovering the evolution of research topics and innovations in scientific archives~\cite{li2021analytic} and understanding trends in public opinion on particular issues~\cite{sha2020dynamic}.

There exist two different cultural methodologies in utilizing statistical modeling to derive insights from data\cite{breiman2001statistical}. Hence, topic models can be studied from two perspectives: Probabilistic Topic Models and Algorithmic Topic Models. Probabilistic Topic Models~\cite{blei2003latent,blei2012probabilistic,steyvers2007probabilistic} assume that each document in a corpus is a mixture of topics, and each topic is a probability distribution over the words in the corpus. While Algorithmic Topic Models~\cite{thompson2020topic,grootendorst2022bertopic,bahrainian-etal-2021-self-supervised,angelov2020top2vec} combine several algorithms and take advantage of numerical optimization techniques to represent topics as a set of weighted words extracted from a cluster of semantically similar documents. This paper tends to study dynamic topic models from the algorithmic point of view. Therefore, it explores Dynamic Topic Models within two classes: Probabilistic Dynamic Topic Models (PDTM) and Algorithmic Dynamic Topic Models (ADTM). 



PDTMs~\cite{teh2004sharing,blei2006dynamic, wang2006topics,wang2012continuous,dieng2019dynamic} assign probabilities to different words and topics over time, allowing it to make inferences about the underlying themes and patterns in the data as they change. On the other hand, ADTMs~\cite{grootendorst2022bertopic,gao2022semantic,eklund2022dynamic} formalize the underlying probability distributions of words and documents into a fixed-length vector representation space using advanced data-mining techniques.

While PDTMs are widely used in various works\cite{greene2017exploring,yao2020tracking}, they become computationally expensive in the face of large archives with very large vocabulary sizes~\cite {bhadury2016scaling}. They lack scalability comparing to ADTMs since they involve sampling from complex posterior distributions~\cite{pmlr-v162-zhang22n}. Besides, PDTMs are not adapted to handling short datasets for applications such as social media monitoring (e.g. Twitter)~\cite{albalawi2020using}, while ADTMs are well-suited for processing short text data with high efficiency~\cite{albalawi2020using,zhang2022meta,feng2022context} since they often take advantage of Large Language Models (LLMs) for feature extraction.

However, the current state-of-the-art PDTMS and ADTMs suffer from certain limitations. One such limitation is that they are constrained to maintaining the same number of topics in each period, often with identical global parameterization. Consequently, there are instances where documents appear in the same global topics despite actually belonging to different topics within a specific time frame. This issue hampers the ability to capture the temporal variations of dynamic topics, including changes in the number of document clusters or the size of each cluster, as topics emerge and fade over time. Moreover, this limitation diminishes the explainability and interpretability of evolving topics within specific periods. This is primarily due to the fact that the representation is derived solely from the content of documents assigned to a set within a given period.
This limitation affects the observation of topic evolution and reduces the quality of topics presented in each period. 

\begin{figure*}[ht]
\centering
\includegraphics[width=0.8\textwidth]{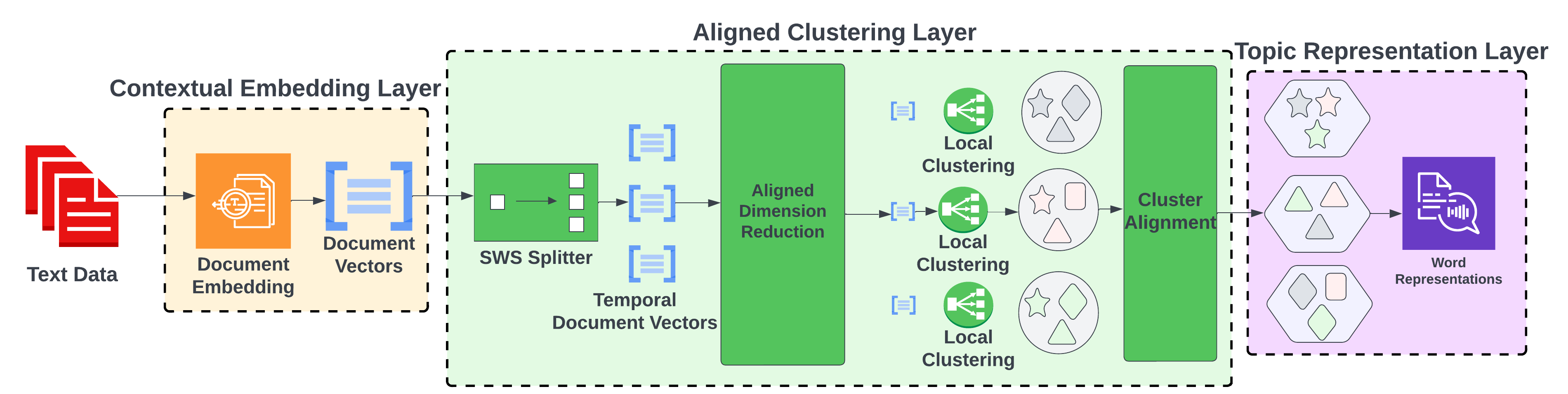}
    \caption{The Architecture of ANTM. \textnormal{The first layer uses advanced pre-trained LLMs to provide a time-aware vector representation for each document. The second layer splits the document vector representations into a set of temporal time frames and performs an overlapping sliding window algorithm for temporal document clustering. Finally, the third layer is responsible for providing word representations for each set of aligned clusters over time.}}
    \label{fig:arc}
\end{figure*}

To address this issue, we introduce ANTM, a novel architecture (as shown in \Cref{fig:arc}) for algorithmic dynamic topic modeling which takes the temporal variations of dynamic topics into account and produces diverse and high-quality topics. 
ANTM discovers evolving topics through an overlapping sliding window algorithm for temporal document clustering. This algorithm called Aligned Clustering is composed of slicing archives into overlapping segments, performing sequential density-based clustering, and aligning similar adjacent document clusters over different periods. 
Aligned clustering allows for the identification of sets of similar topics that span multiple time periods and yet are different enough to show some sort of evolution.
ANTM takes advantage of advanced pre-trained Large Language Models (LLMs)~\cite{chen2021evaluating} such as Data2Vec~\cite{baevski2022data2vec}, which predicts latent representations of the documents in a self-supervised manner using a standard transformer architecture. 
%
%
Our experiments on four dataset show a considerable improvement in terms of coherence and diversity score over existing PDTMs and ADTMs.
Besides, ANTM can perform effectively on short text applications such as Twitter as shown in \Cref{fig:musk}.

\textbf{Contributions.} We propose ANTM, a new family of dynamic topic models that effectively capture topic evolution using advanced algorithms. Second, we conduct a comprehensive analysis to evaluate the interpretability of the evolving topics generated by ANTM compared with state-of-the-art dynamic topic models, including recent advancements such as DETM \citep{dieng2019dynamic} and BERTopic \citep{grootendorst2022bertopic}, as well as a traditional probabilistic dynamic topic model, DTM \citep{blei2006dynamic}. Third, we will show ANTM outperforms existing dynamic topic models in terms of coherence and diversity, particularly excelling in handling short text datasets.

\textbf{Organization.}The rest of the paper is organized as follows. The state-of-the-art is investigated in \Cref{sec:related}. ANTM is introduced in \Cref{sec:method} and the experiments are presented in \Cref{sec:experiment}. Finally, we will conclude the paper in \Cref{sec:conclusion}.

\section{Related work}
\label{sec:related}



There are various assumptions on the definition of topic evolution that statistically characterize dynamic topic models. DTM (Dynamic Topic Model)~\cite{blei2006dynamic} is among the first PDTMs\cite{teh2004sharing} to incorporate temporal components to study topic evolution. DTM is a variant of the generative probabilistic model called Latent Dirichlet Allocation (LDA)~\cite{blei2003latent}. DTM creates topics that consist not only of a set of words but also of a timestamp or time range. It uses a Bayesian approach to model the change in the relative proportions of topics within documents over time. DTM is extended by several new models\cite{iwata2010online,bahrainian2017modeling,wang2012continuous,wang2006topics,ren2008dynamic}. For instance, authors in \cite{wang2006topics} stated that time is inherently continuous and therefore they characterize their proposed model with a continuous distribution over timestamps. One other example is the Discrete-Time Dynamic Topic Model (dDTM) \cite{bahrainian2017modeling}, which requires the data to have discrete time intervals, while the Continuous-Time Dynamic Topic Model (cDTM)~\cite{wang2012continuous} can handle any data point in time, regardless of the time resolution.

Although DTM has been an innovative tool for studying topic evolution at the time, it has several limitations. The first of these is scalability, where the DTM becomes computationally expensive and time-consuming when dealing with very large data sets. In addition, DTM cannot fully capture the diversity of topics over time because it assumes that topics remain relatively stable over time. Therefore, topics are consistent across the dataset and one may not be able to detect fine-grained changes in topics or subtle variations in how topics are represented in different periods. In addition, the number of topics must be known before modeling. There are also several generative probabilistic topic models~\cite{wei2007dynamic,10.1145/1835804.1835889,10.1145/2872427.2883046,zosa2019multilingual} that attempt to overcome these problems. For example, DTM is combined with word embeddings in Dynamic Embedded Topic Models (DETM)~\cite{dieng2019dynamic,dieng2020topic} to improve the performance of topic models by providing a more informative representation of words. However, it is still very time-consuming and we will show that its topic representations are of lower quality in terms of diversity and coherence compared to ADTMs. As another example, to reduce the computational complexity and the number of variational parameters for handling large vocabularies, an amortized variational inference method based on DTM is developed in~\cite{pmlr-v180-tomasi22a}. There are also efforts in the development of scalable PDTMs, such as~\cite{pmlr-v84-jahnichen18a}, which extends the class of tractable priors from Wiener processes to the more general class of Gaussian processes. This approach allows the model to be applied to large collections of text and to explore topics that evolve smoothly over a long period.

In spite of all these efforts to address the problems of PDTMs, they have still been shown to have less coherence and diversity in topic representation compared to ADTMs~\cite{grootendorst2022bertopic,gao2022semantic,eklund2022dynamic}. ADTMs formalize the underlying temporal probability distributions of words and documents into a fixed-length vector representation space through advances in neural networks. While preserving important words in the topic descriptions, these methods generate dense clusters of interpretable documents. Among the most widely used ADTMs are BERTopic\cite{grootendorst2022bertopic}, which, unlike PDTMs, allows context to be embedded in the generated topics through the use of Large Language Models (LLMs) that generate document representations. These embedded vectors are then divided into a set of clusters that describe a set of documents that describe semantically similar topics across an archive. BERTopic combines BERT and c-TF-IDF~\cite{grootendorst2022bertopic} to generate dense clusters of interpretable topics while preserving important words in the topic descriptions. By computing the c-TF-IDF topic representation in each time window, BERTopic allows dynamic topic modeling. Although BERTopic can handle large and short text datasets (due to LLMs), it still creates static clusters over the whole dataset and only represents them dynamically. This process ignores the temporal structure of dynamic topic models in each period. Furthermore, it limits the topic model to have the same number of topics for each period. In fact, we will show that grouping documents statically across the corpus and representing them dynamically based on global document clusters affects the observation of topic evolution and reduces the quality of topics represented in each period. The method proposed in this paper is fast and more scalable than PDTMs. In addition, it can take into account the limitations discussed in BERTopic and can produce higher-quality word representations which allows for more detailed exploration of topic evolution.

\section{ANTM}
\label{sec:method}

As shown in \Cref{fig:arc}, ANTM consists of three layers: The first layer uses advanced pre-trained Transformer-based LLMs (e.g. Data2Vec) to provide a time-aware vector representation for each document, corresponding to its content as it relates to other documents. The second layer splits the document vector representations into a set of temporal time frames and performs an overlapping sliding window algorithm (based on AlignedUMAP\cite{alignedumap} and a hierarchical density-based clustering algorithm called HDBSCAN) for temporal document clustering. Finally, the third layer is responsible for providing word representations for each set of aligned clusters over time. The class-based TF-IDF method is used in this layer.

\subsection{Contextual Embedding Layer (\CEL)}
The Contextual Embedding Layer (\CEL) is responsible for providing a vector representation $y$ for the document $d$ in the corpus $D$. More formally, the embedding $y$ for a document $d$ is a mapping $\CEL: d\mapsto y \in \Re^z$, capturing contextual and semantic information from the corpus $D$. The document embedding represents words and documents in a low-dimensional feature vector space, where the embedding dimension $z$ is expected to be much smaller than the size of the vocabulary (i.e., the number of unique words in $D$). $\CEL$  takes advantage of pre-trained transformer-based LLMs (e.g., Data2Vec, GPT4) trained on a large corpus of text data to compute time-aware vector representation for each document. Transformer-based models capture the meaning and context of a document by using an attention mechanism to consider the context of words in a document. Given an input document $d$, the number $u$ of tokens in $d$, and $\textbf{h}_j$  the hidden state vector of the $j$-th token in $d$, the embedding $y = \CEL(d)$ of the whole document $d$ is then obtained by taking the mean of all the hidden state vectors of the tokens in $d$.
The attention mechanism of Transformers allows the model to focus on different parts of the input text, weighting the importance of each word or phrase in the context of the whole document. This empowers $\CEL$ to capture the meaning of the document as well as its temporal aspect.

\subsection{Aligned Clustering Layer (ACL)}

While we agree that time is inherently continuous, this layer is motivated by the belief that topic evolution occurs in cycles as explained by~\cite{kuhn2012structure}. Therefore, we discretize these cycles by time frames with intersections similar to \cite{blei2006dynamic,hu2015modeling}, but we use time-aware algorithms to account for the continuity of time and to model evolving topics that emerge and fade over time. After performing transformer-based document embedding on the corpus $D$, we can obtain a dynamic vector representation of $D$ by applying a Sliding Window Segmentation ($\SWS$) process that divides $D$ into a series of $n$ overlapping time frames $\{W^1,\dots, W^n\}$.  We then denote the set of documents published in the time frame $W^t$ by $D^t\subseteq D$ and the set of embeddings of $D^t$ by $Y^t= \{\CEL(d) | d \in D^t \}$. This layer aims to discover evolving document clusters by sequentially grouping similar documents. This procedure is called Partitioned Clustering ($\PC$). 

 {\definition Let $D^t$ be a set of documents in the $t$-th time frame of the document corpus $D$ and let $Y^t$ denote the set of embeddings of $D^t$. Partitioned Clustering $\PC: D^t \mapsto \{D_i^t\}$ clusters the documents $D^t$ by their embeddings $Y_t$ and returns a set of \textit{local} document clusters $D_i^t\subseteq D$ for $i=1$  to  $k_t$ where \emph{all document embeddings in $Y_i^t$ are similar}.
}

\begin{figure*}[htbp]
  \centering
  \begin{subfigure}[b]{0.22\textwidth}
    \includegraphics[width=0.9\textwidth]{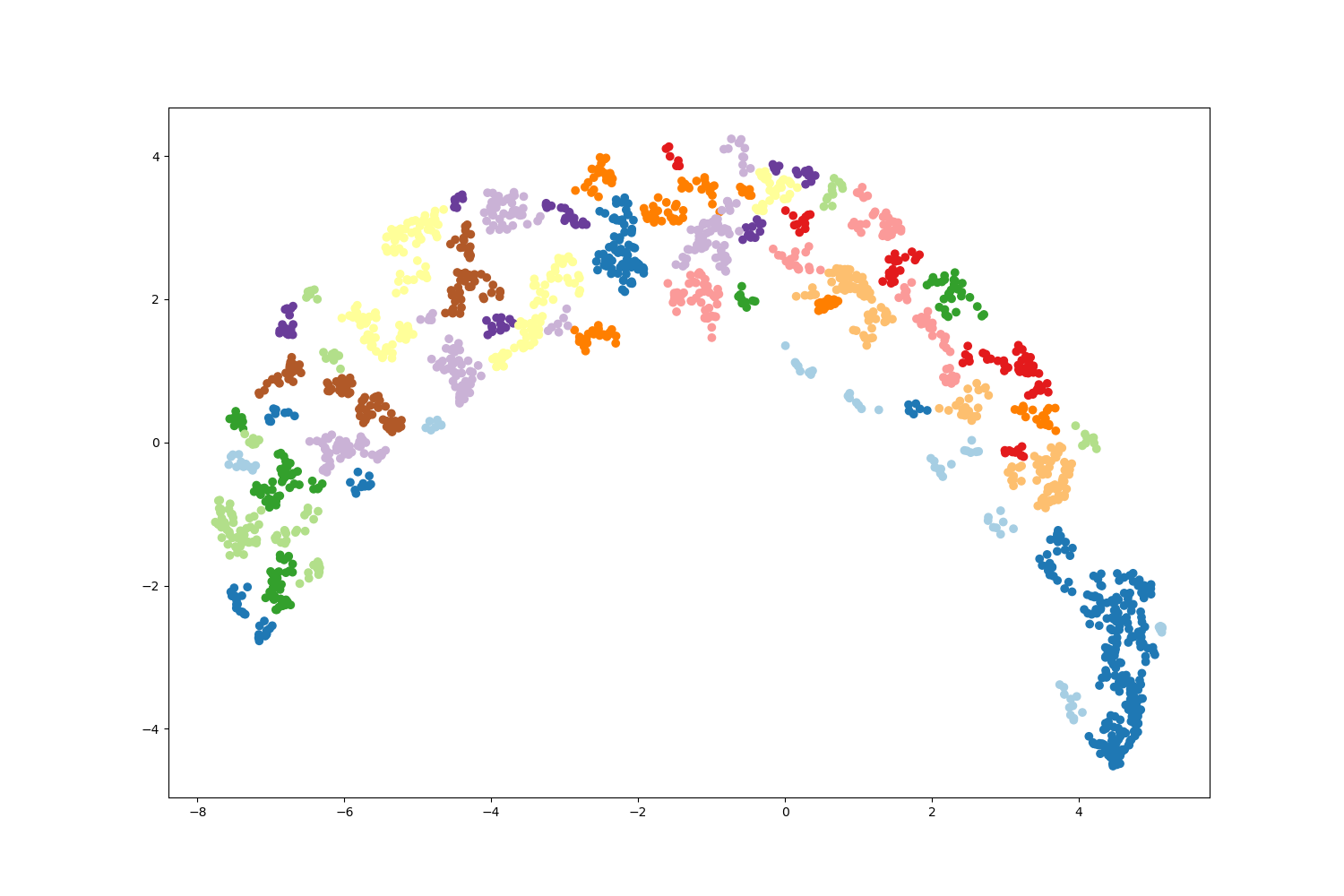}
    \caption{2006-2008}
  \end{subfigure}
  \hfill
  \begin{subfigure}[b]{0.22\textwidth}
    \includegraphics[width=0.9\textwidth]{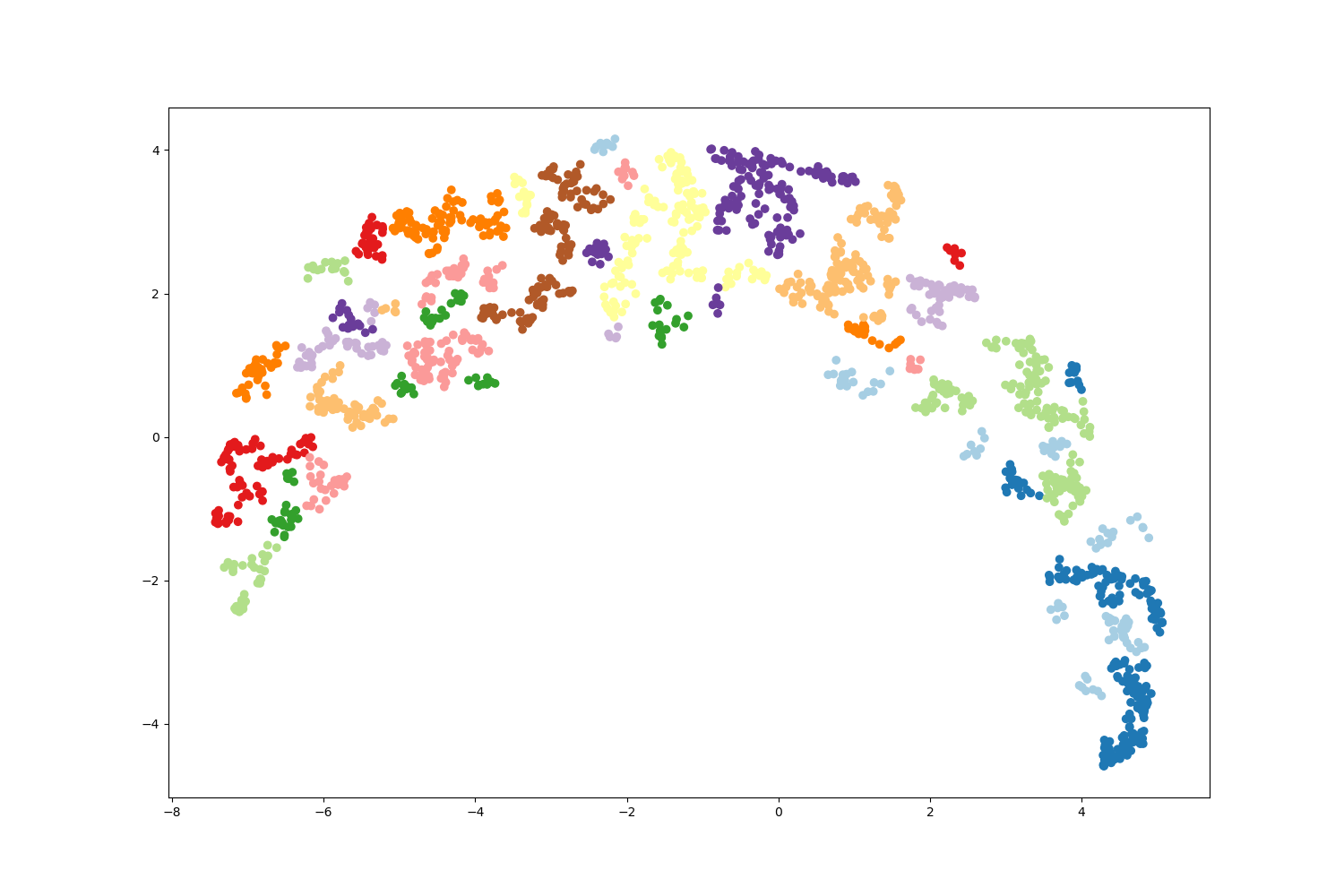}
    \caption{2008-2010}
  \end{subfigure}
  \hfill
  \begin{subfigure}[b]{0.22\textwidth}
    \includegraphics[width=0.9\textwidth]{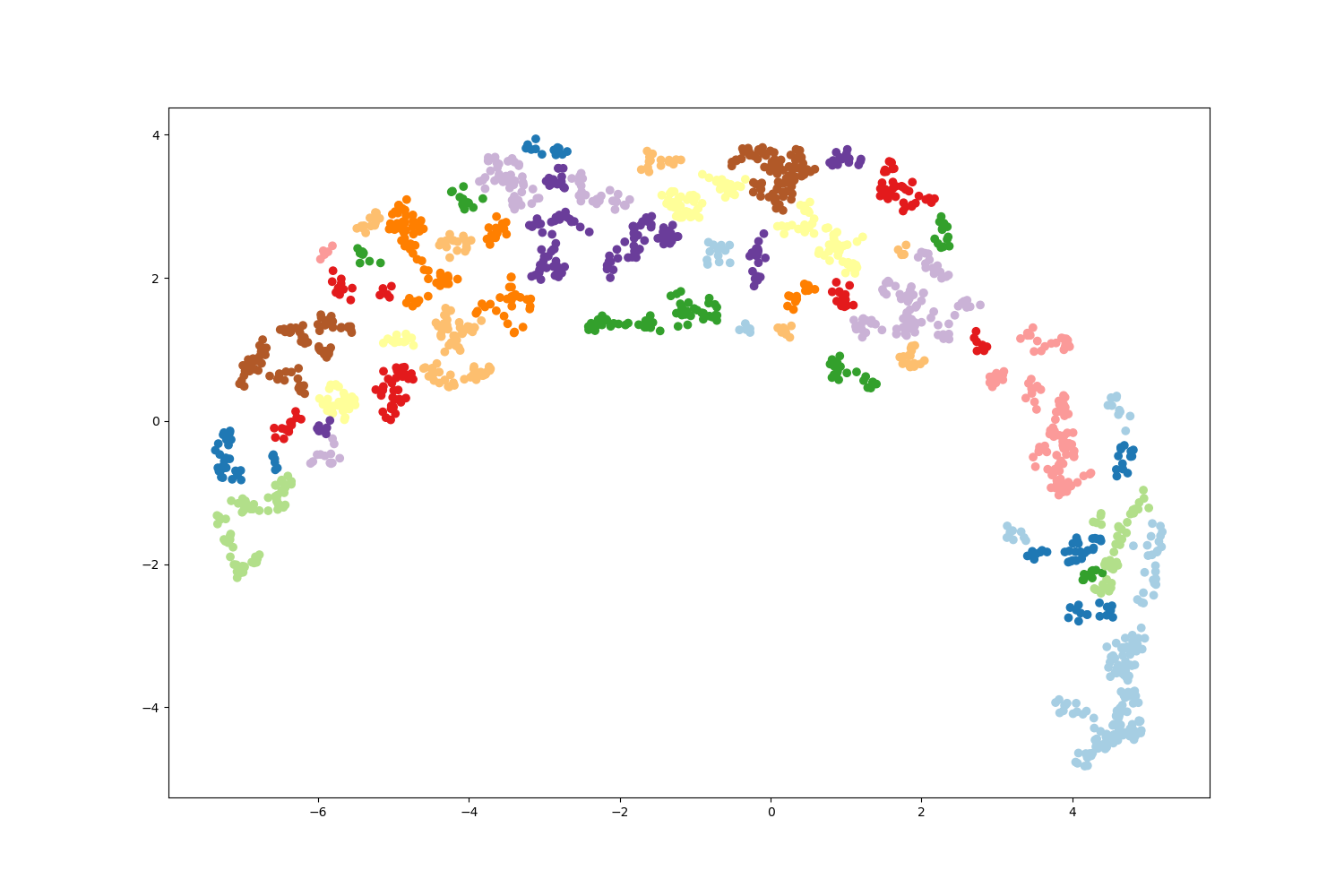}
    \caption{2010-2012}
  \end{subfigure}
  \hfill
  \begin{subfigure}[b]{0.22\textwidth}
    \includegraphics[width=0.9\textwidth]{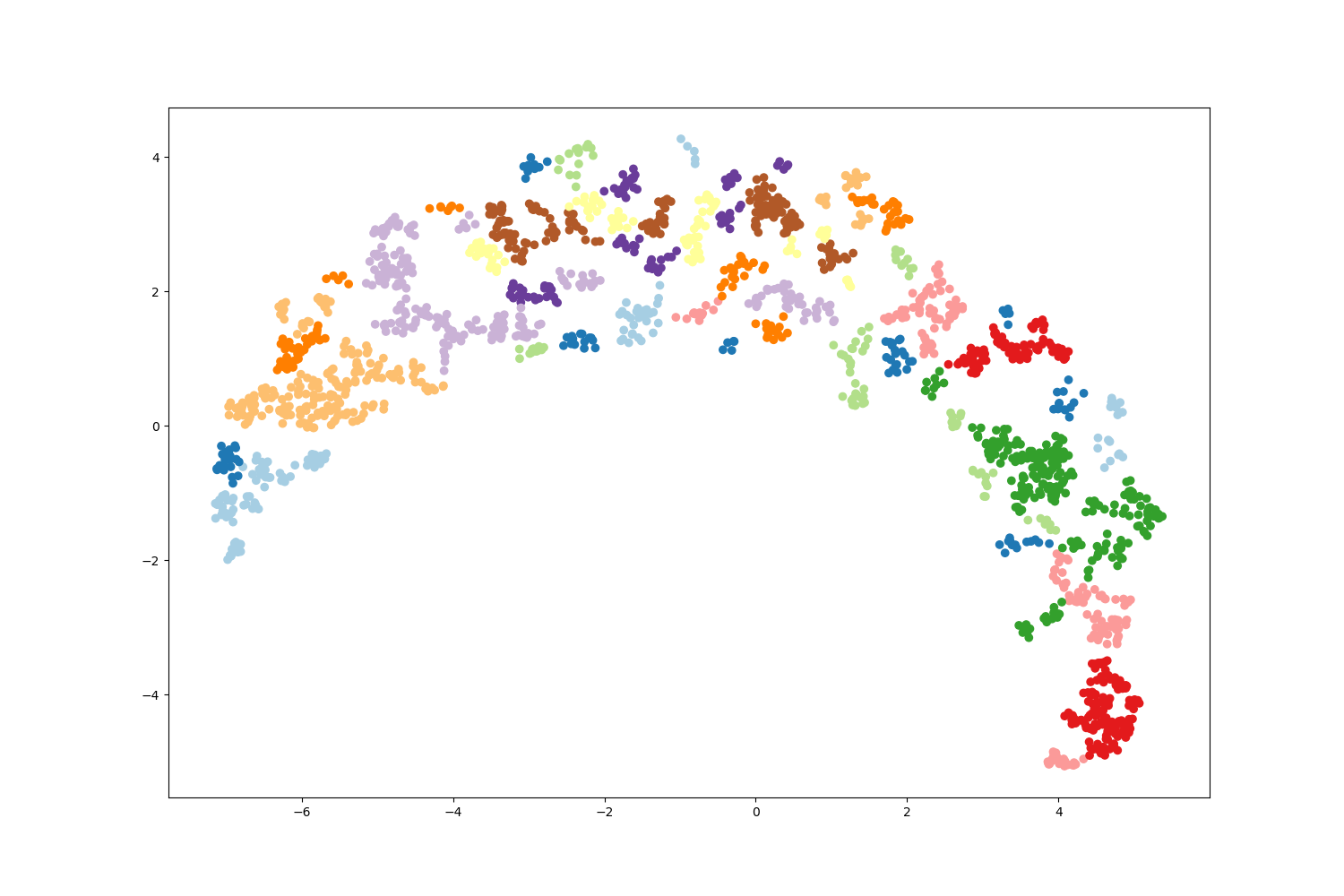}
    \caption{2012-2014}
  \end{subfigure}
  \hfill
  \begin{subfigure}[b]{0.22\textwidth}
    \includegraphics[width=0.9\textwidth]{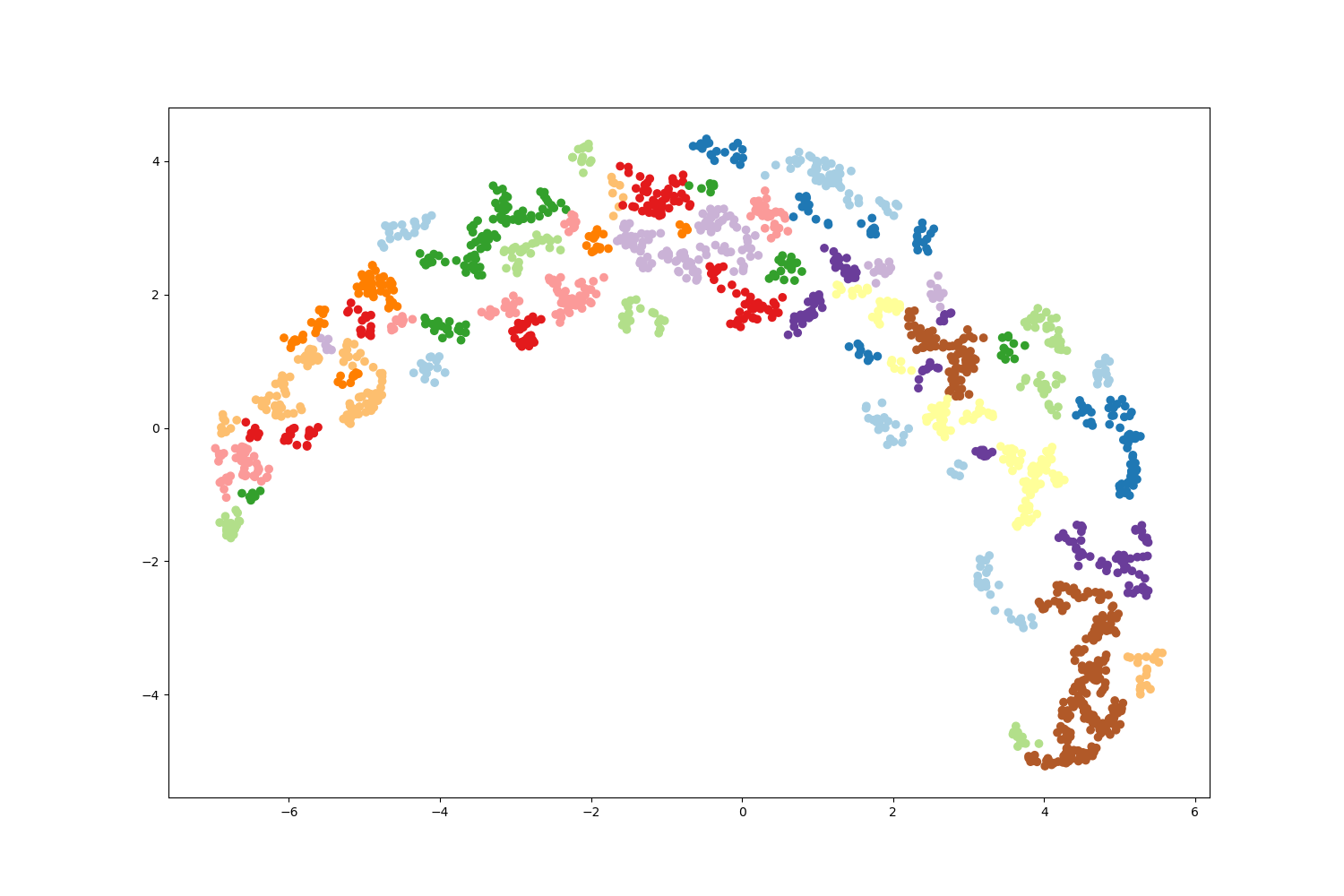}
    \caption{2014-2016}
  \end{subfigure}
  \hfill
  \begin{subfigure}[b]{0.22\textwidth}
    \includegraphics[width=0.9\textwidth]{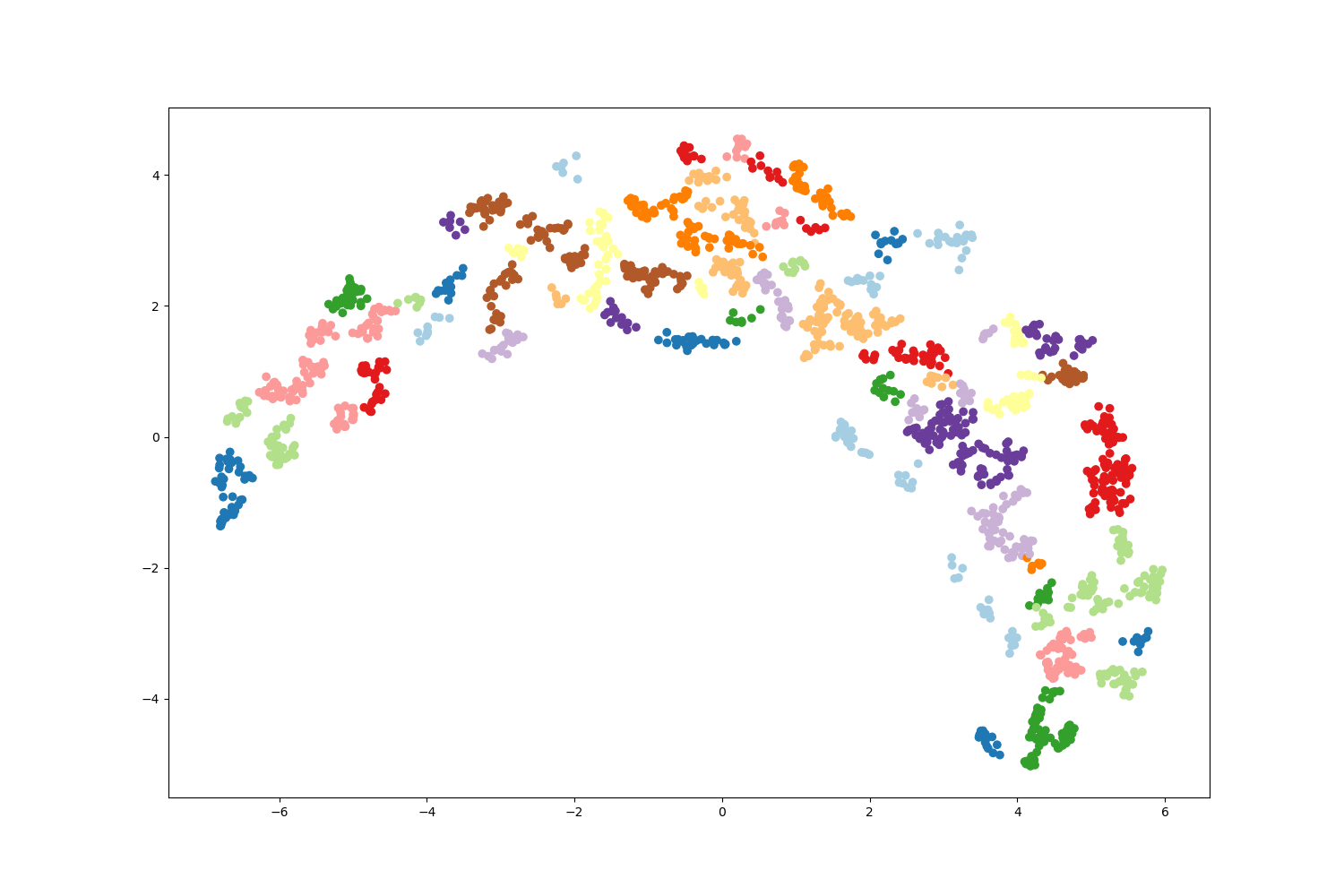}
    \caption{2016-2018}
  \end{subfigure}
  \hfill
  \begin{subfigure}[b]{0.22\textwidth}
    \includegraphics[width=0.9\textwidth]{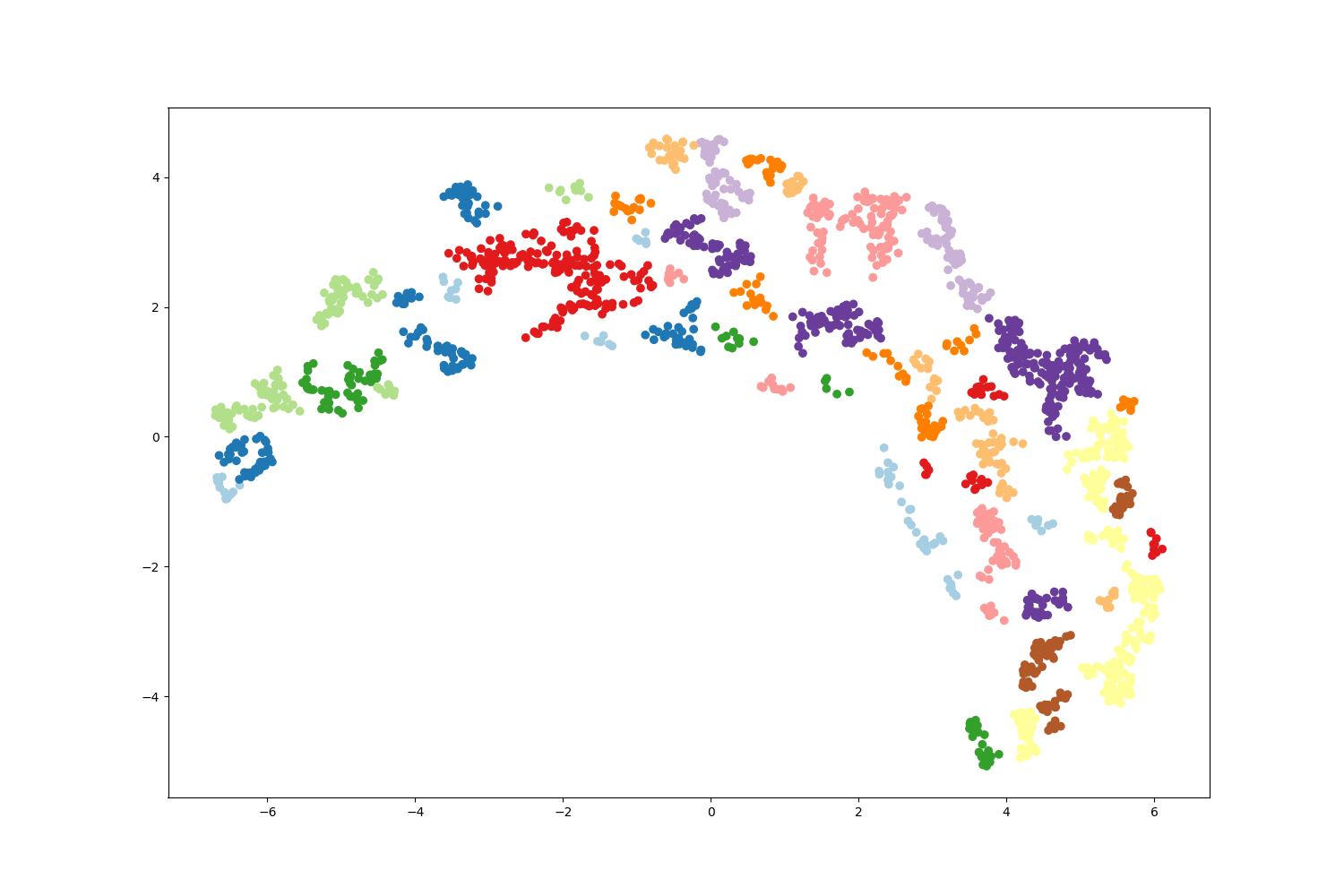}
    \caption{2018-2020}
  \end{subfigure}
  \hfill
  \begin{subfigure}[b]{0.22\textwidth}
    \includegraphics[width=0.9\textwidth]{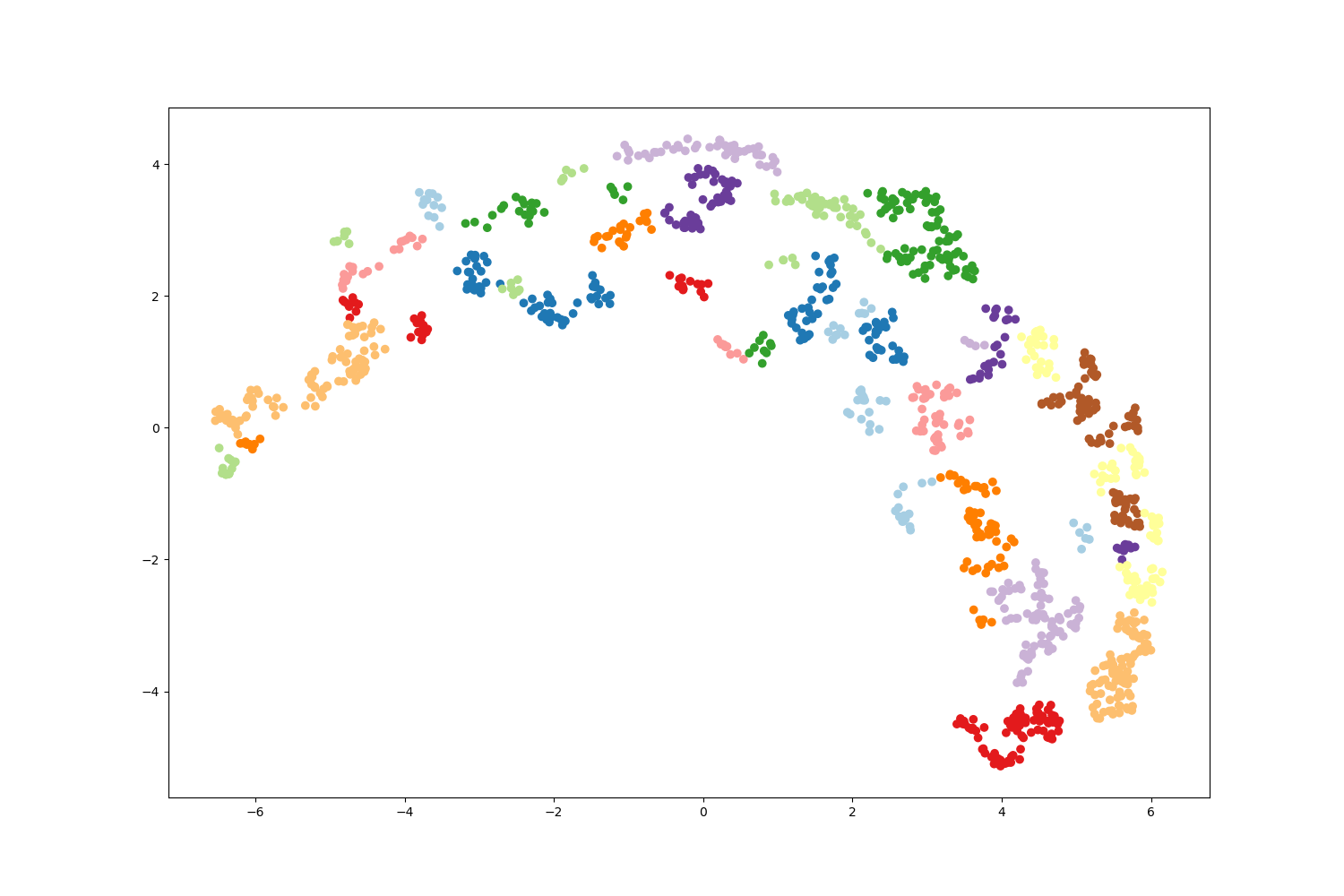}
    \caption{2020-2022}
  \end{subfigure}
  \caption{Partitioned Clusters. \textnormal{The documents from the arXiv dataset are embedded using Data2Vec and sequentially clustered in each time frame. These clusters are consecutively aligned to create evolving topics as described in \Cref{fig:3d}(a).}}
    \label{fig:clusterD2V}
\end{figure*}

\begin{figure*}[htbp]
   \centering
   \begin{subfigure}[b]{0.17\textwidth}
     \includegraphics[width=\textwidth]{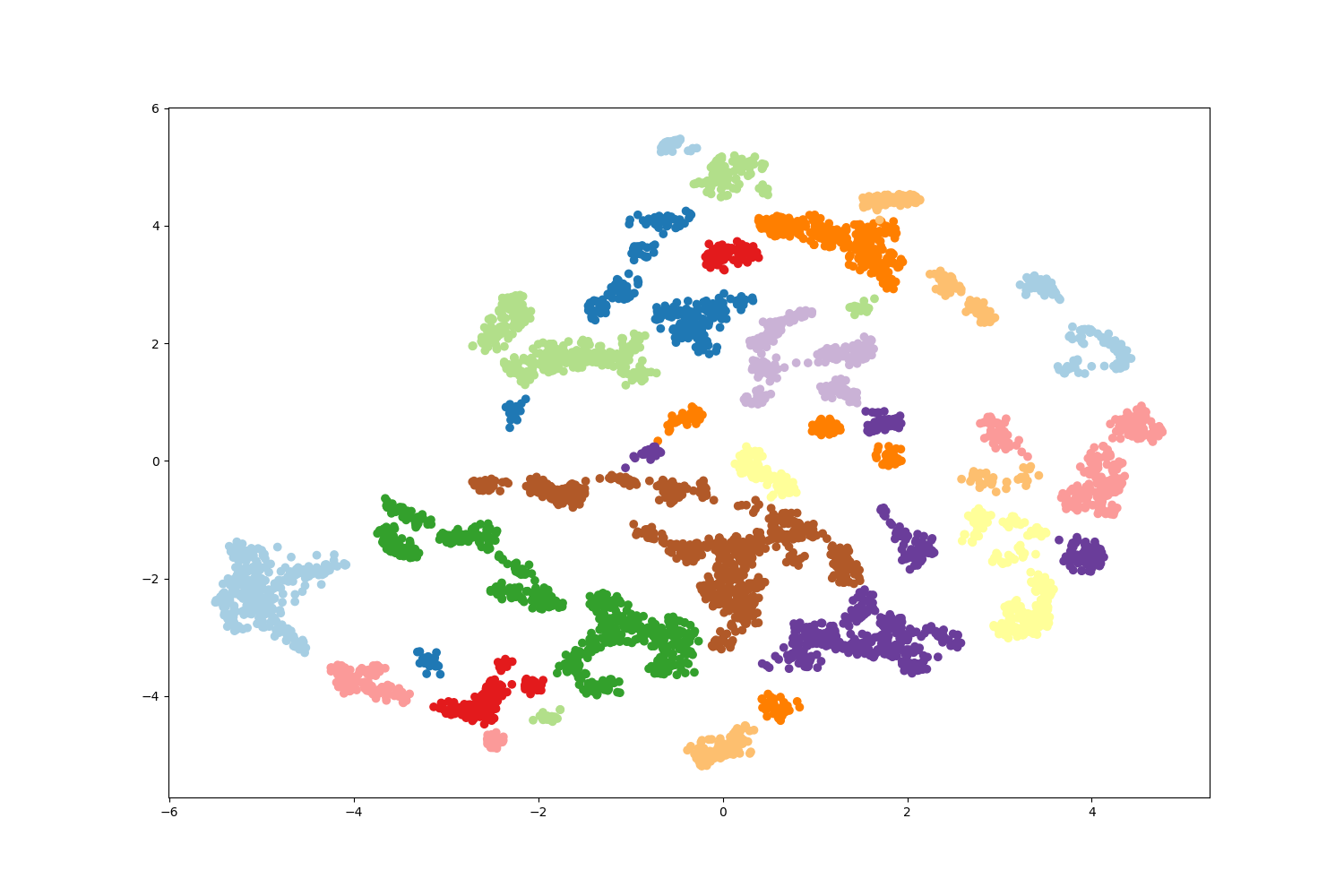}
     \caption{2000-2002}
   \end{subfigure}
   \hfill
   \begin{subfigure}[b]{0.17\textwidth}
     \includegraphics[width=\textwidth]{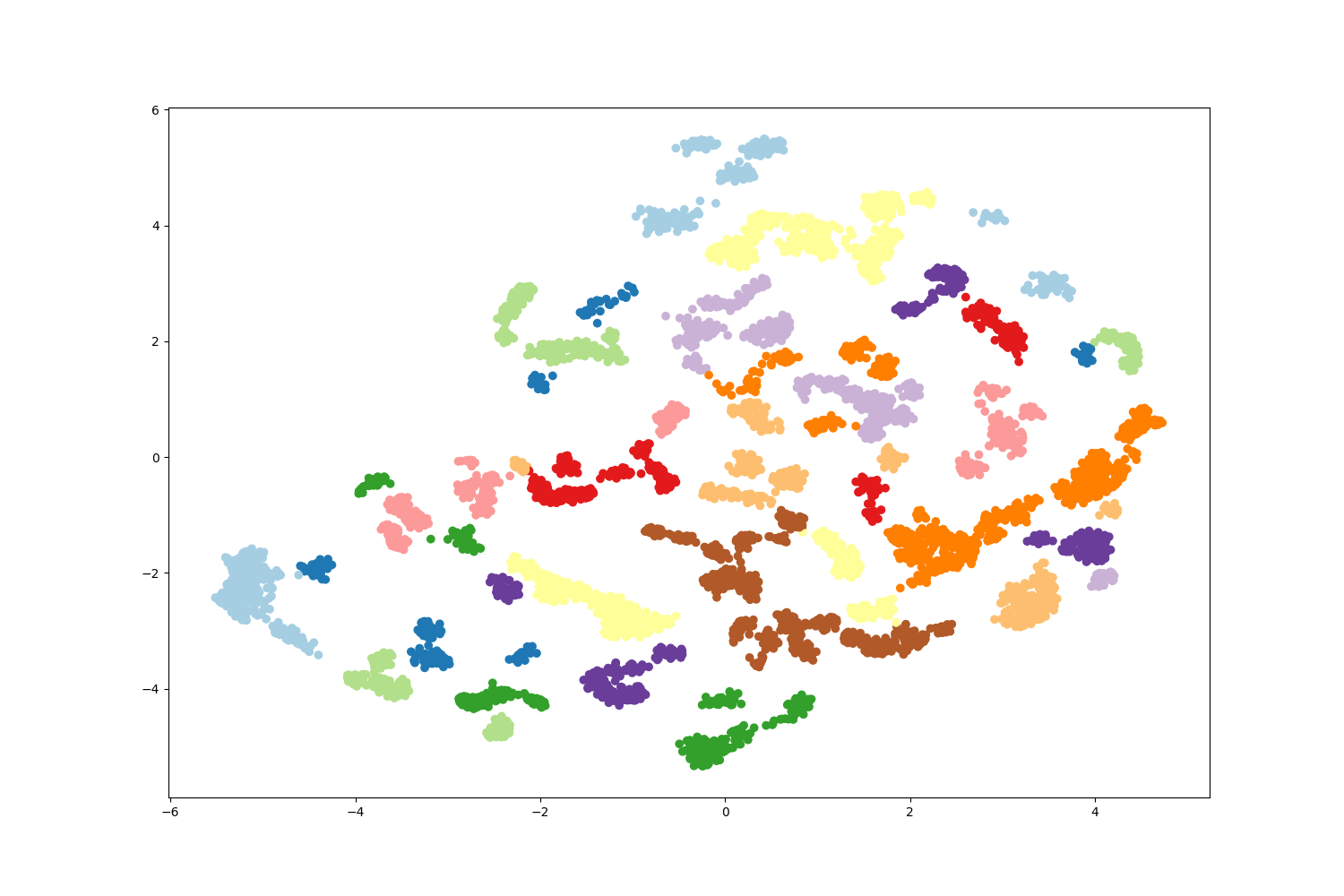}
     \caption{2002-2004}
   \end{subfigure}
   \hfill
   \begin{subfigure}[b]{0.17\textwidth}
     \includegraphics[width=\textwidth]{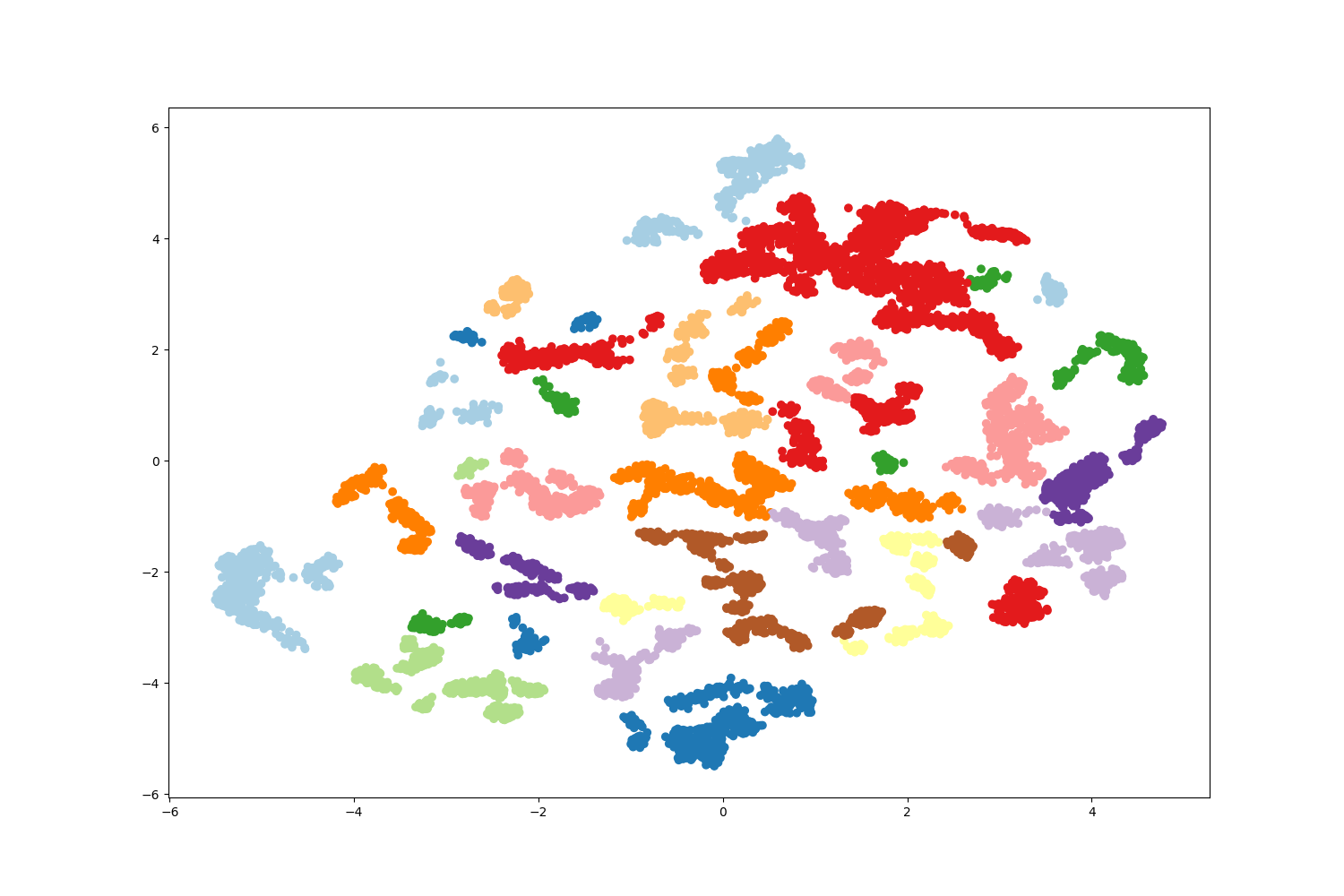}
     \caption{2004-2006}
   \end{subfigure}
   \hfill
   \begin{subfigure}[b]{0.17\textwidth}
     \includegraphics[width=\textwidth]{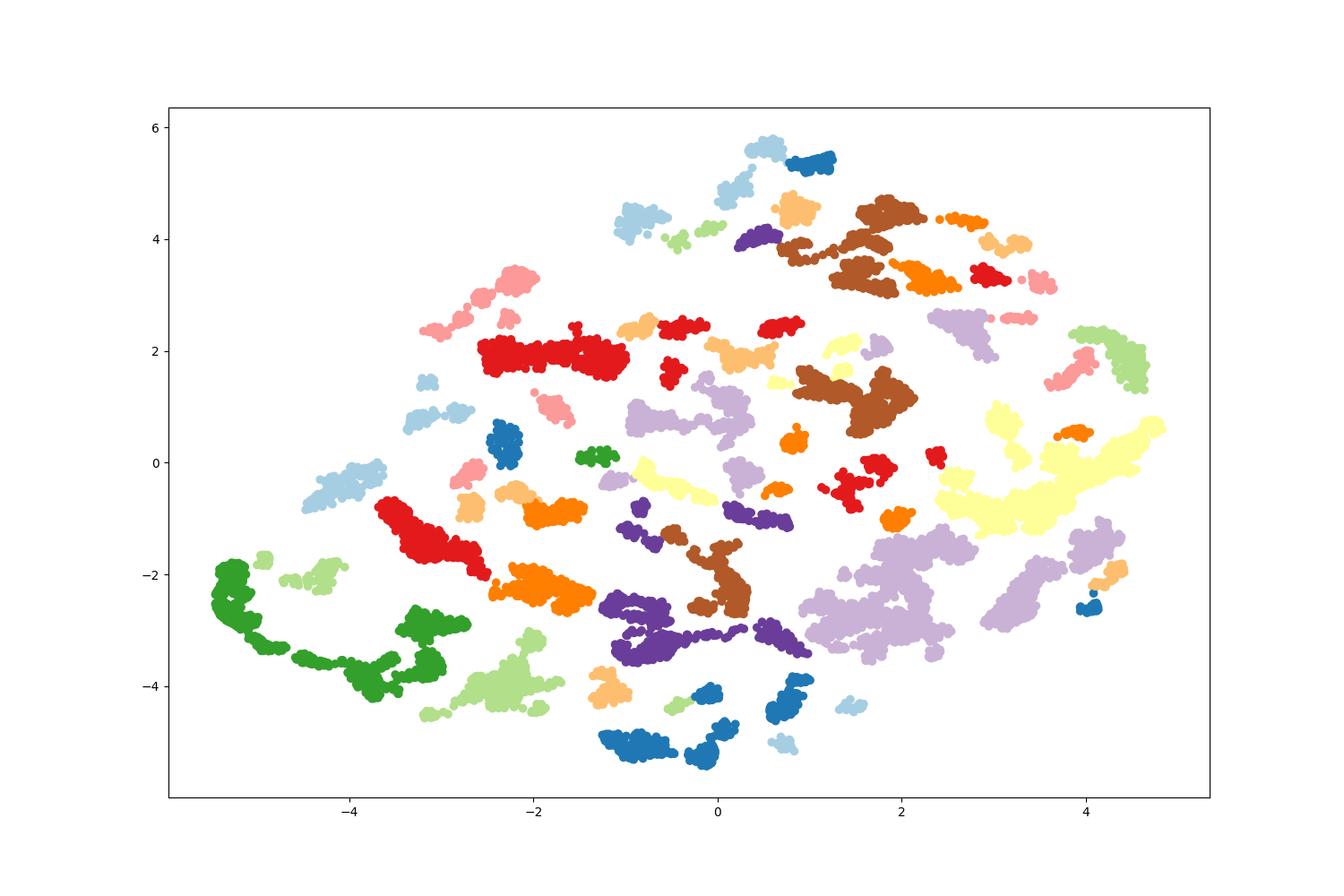}
     \caption{2006-2008}
   \end{subfigure}
   \hfill
   \begin{subfigure}[b]{0.17\textwidth}
     \includegraphics[width=\textwidth]{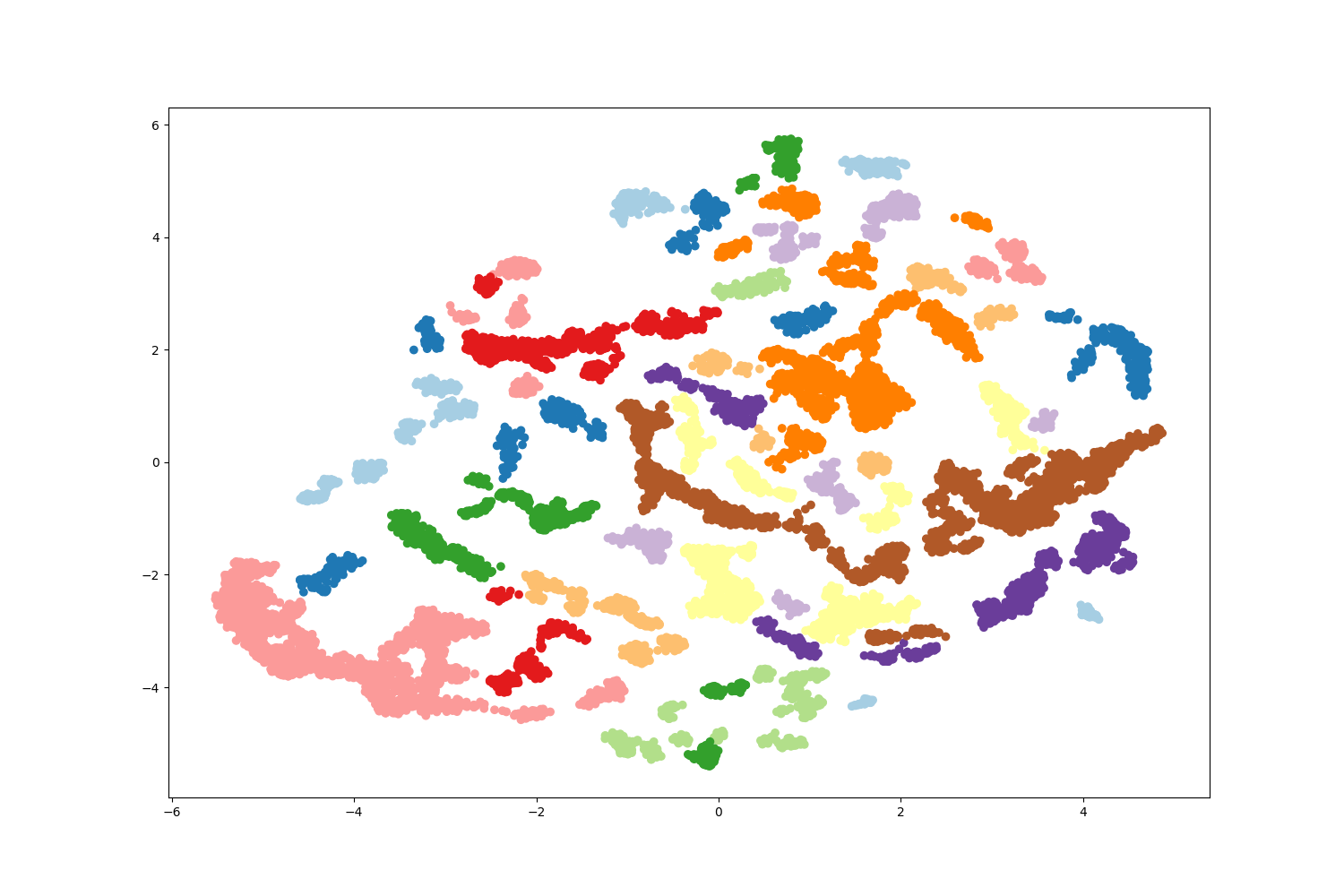}
     \caption{2008-2010}
   \end{subfigure}
   \hfill
   \begin{subfigure}[b]{0.17\textwidth}
     \includegraphics[width=\textwidth]{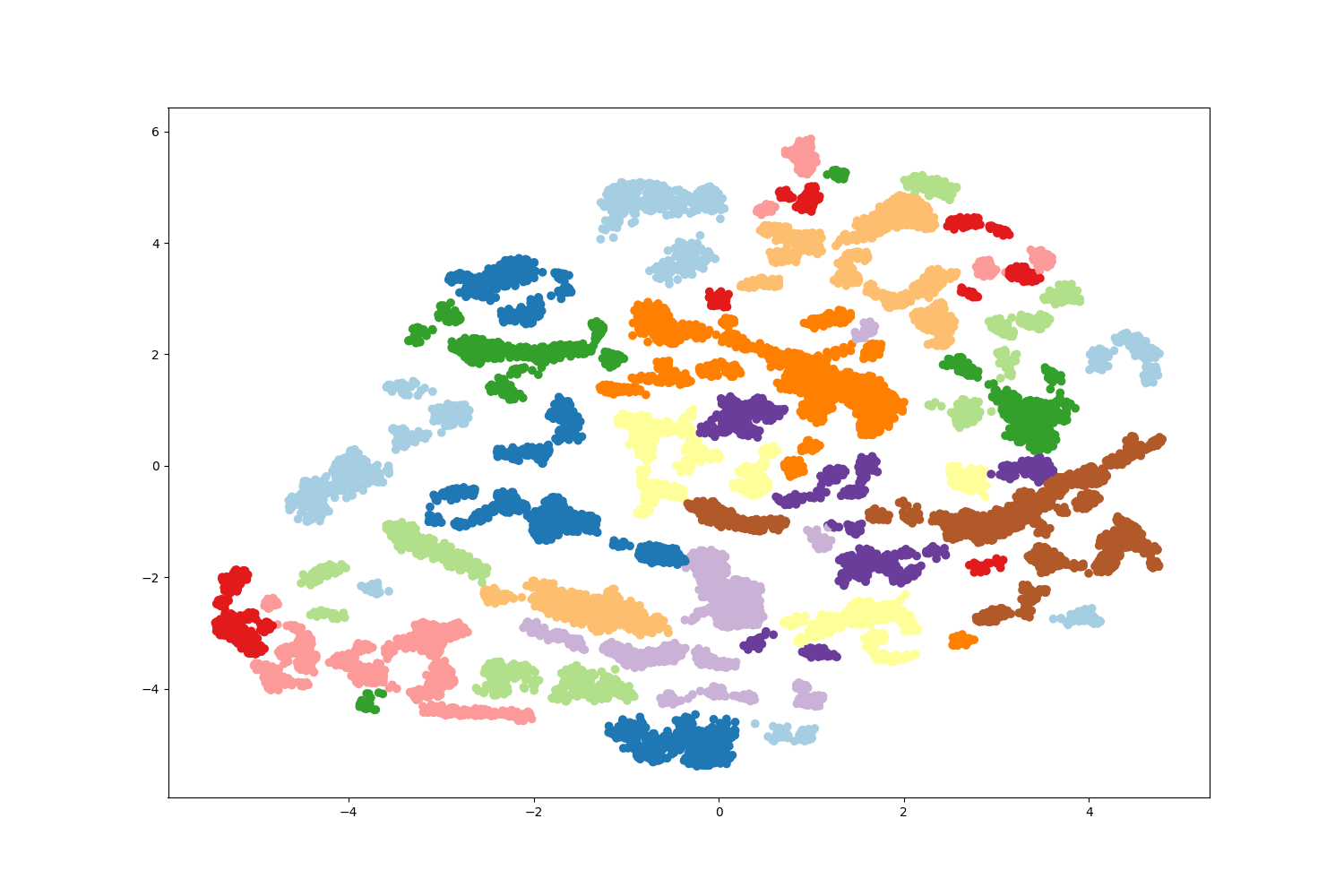}
     \caption{2010-2012}
   \end{subfigure}
   \hfill
   \begin{subfigure}[b]{0.17\textwidth}
     \includegraphics[width=\textwidth]{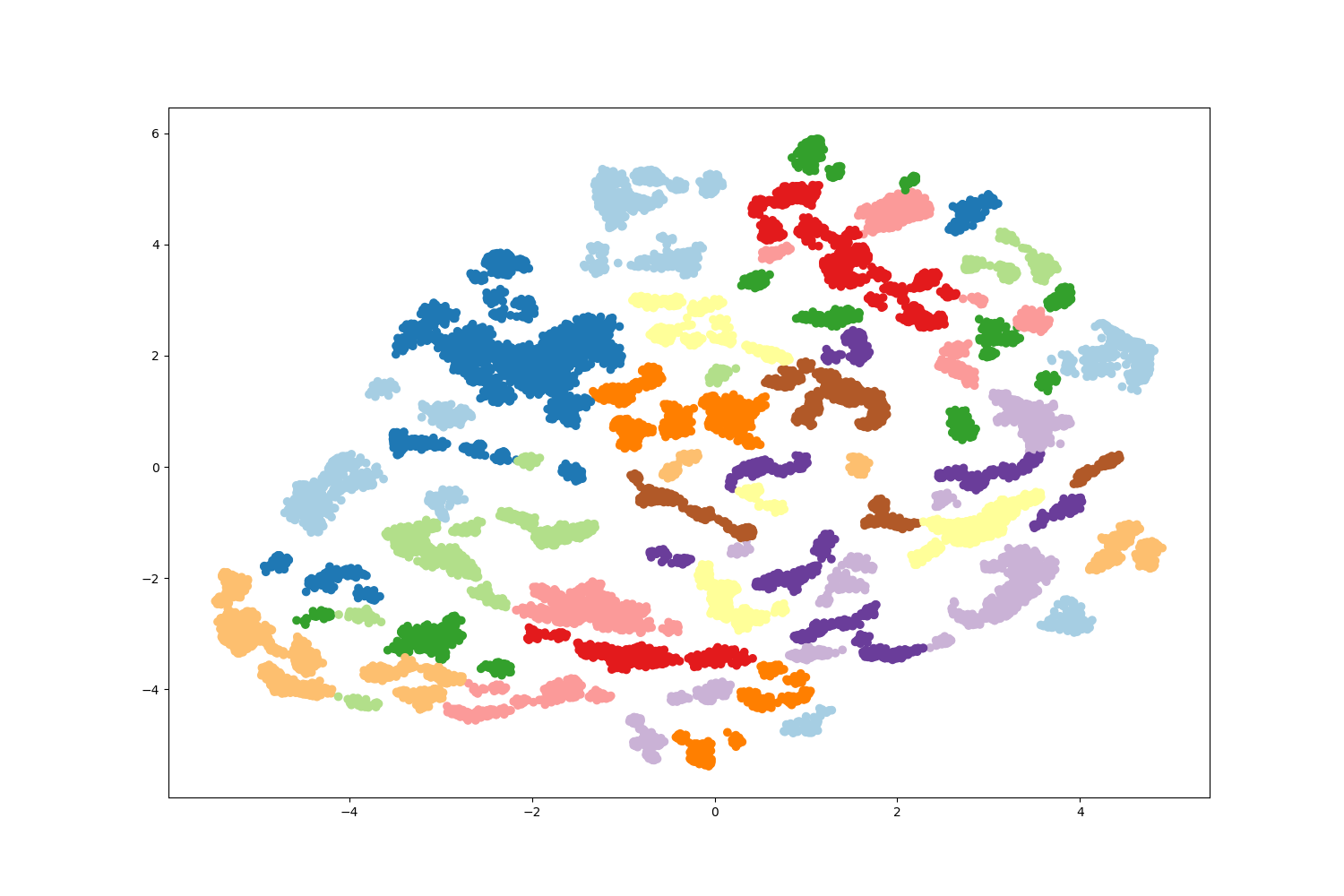}
     \caption{2012-2014}
   \end{subfigure}
   \hfill
   \begin{subfigure}[b]{0.17\textwidth}
     \includegraphics[width=\textwidth]{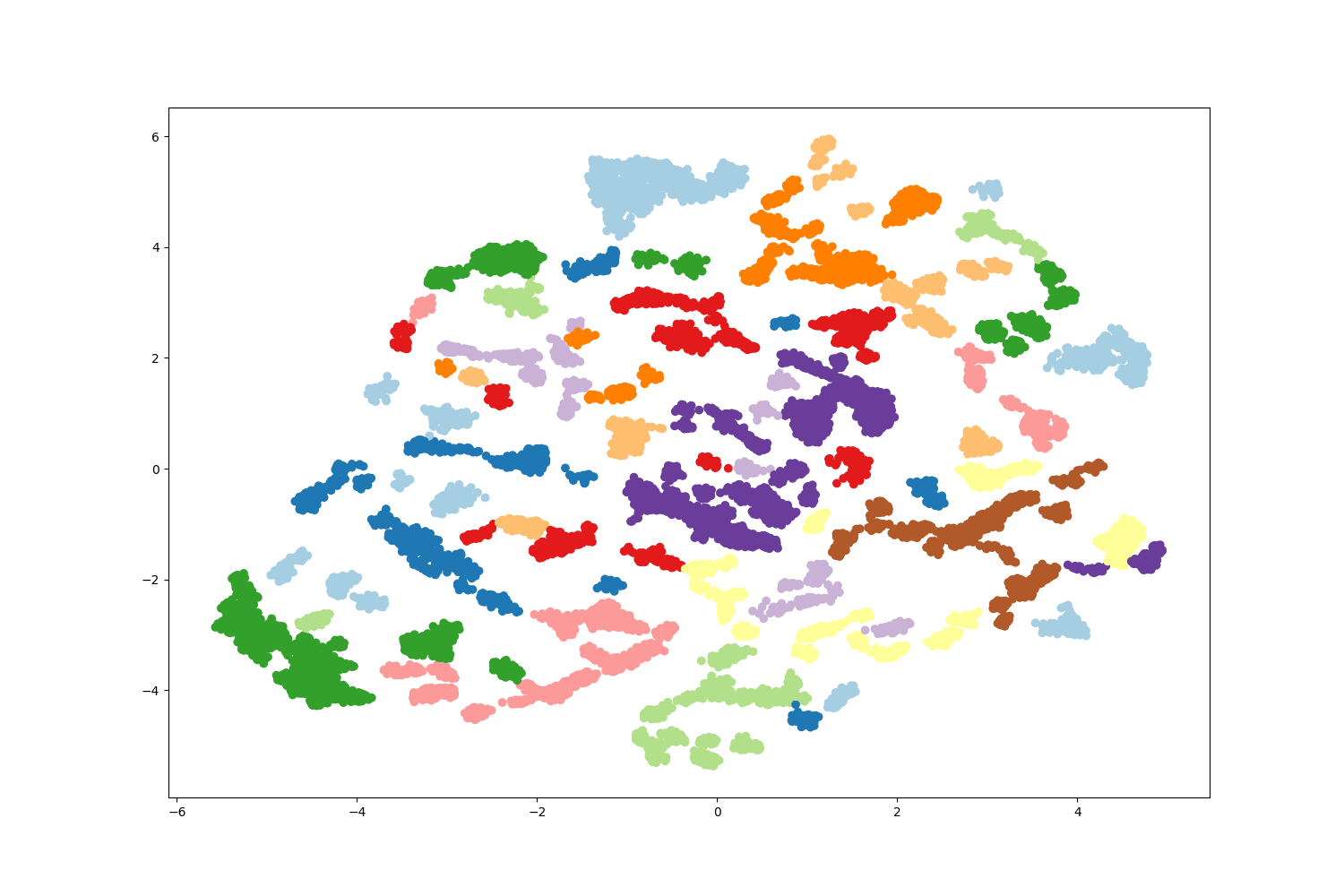}
     \caption{2014-2016}
   \end{subfigure}
   \hfill
   \begin{subfigure}[b]{0.17\textwidth}
     \includegraphics[width=\textwidth]{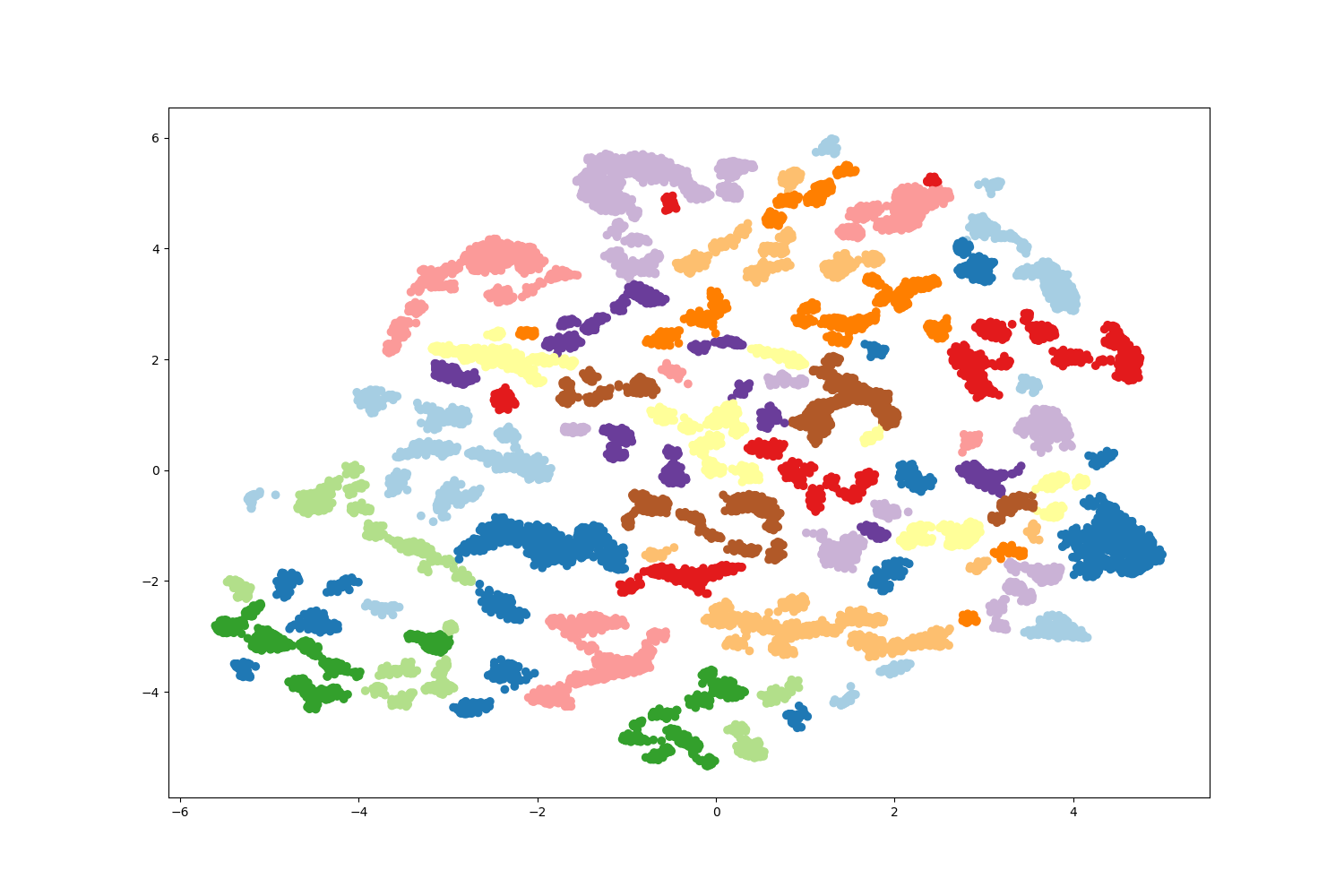}
     \caption{2016-2018}
   \end{subfigure}
   \hfill
   \begin{subfigure}[b]{0.17\textwidth}
     \includegraphics[width=\textwidth]{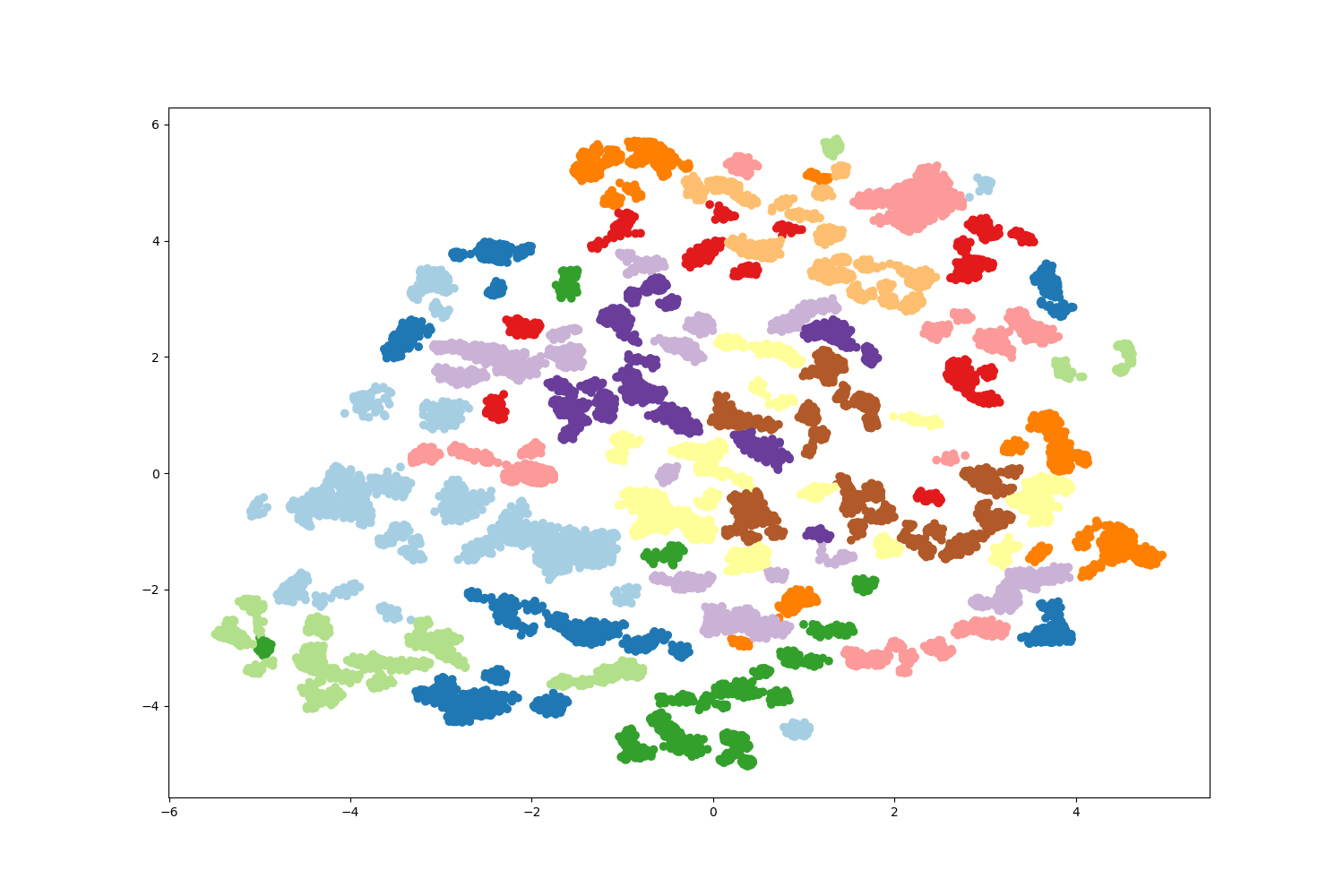}
     \caption{2018-2020}
   \end{subfigure}
 \caption{Partitioned Clusters. \textnormal{The documents from DBLP dataset are embedding using BERT and sequentially clustered in each time frames. These clusters are consecutively aligned to create evolving topics as described in \Cref{fig:3d}(b).}}
     \label{fig:bertDBLP}
 \end{figure*}

As shown in \Cref{fig:clusterD2V,fig:bertDBLP}, clustering documents by the similarity of their embeddings reveals evolution patterns and trends in the data that may not be apparent when documents are grouped based on their content alone. 

\begin{figure*}[htbp]
  \centering
  \begin{subfigure}[b]{0.4\textwidth}
   \centering
    \includegraphics[width=\textwidth]{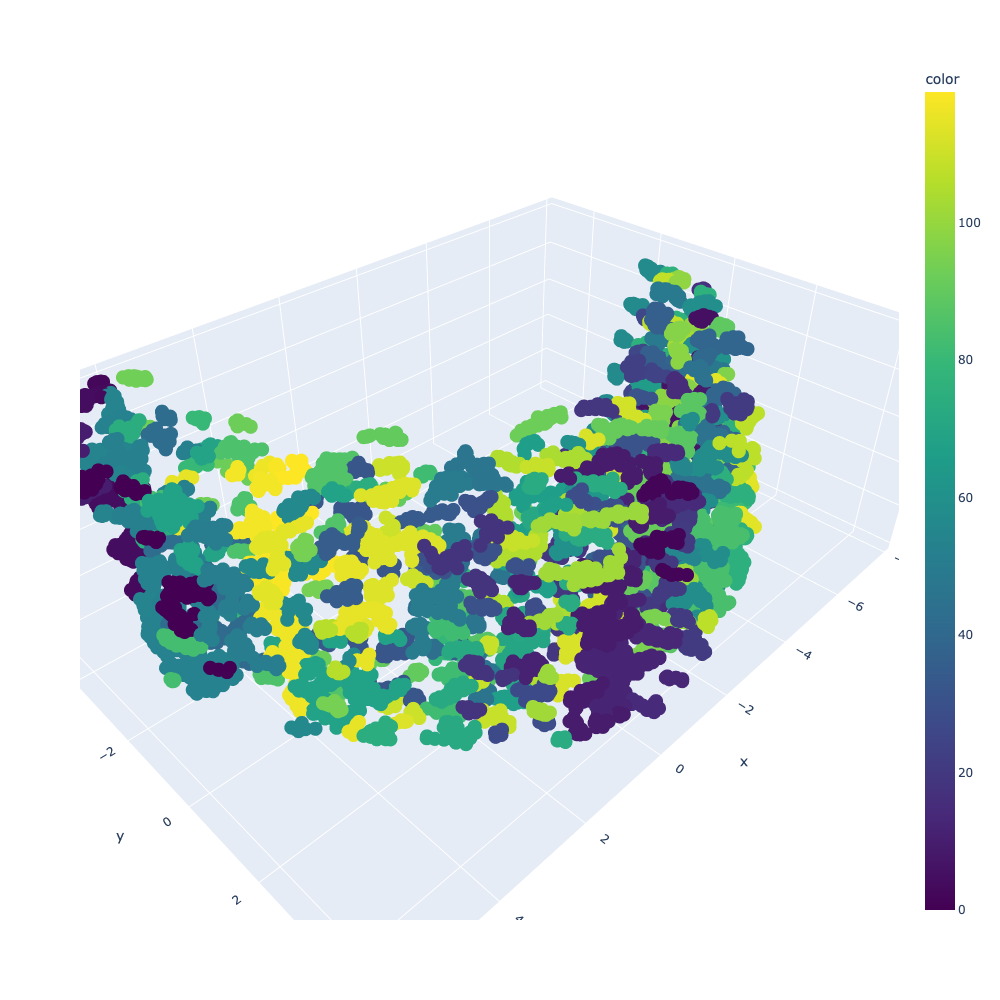}
    \caption{Data2Vec | arXiv}
  \end{subfigure}
   \hfill
  \begin{subfigure}[b]{0.4\textwidth}
   \centering
    \includegraphics[width=\textwidth]{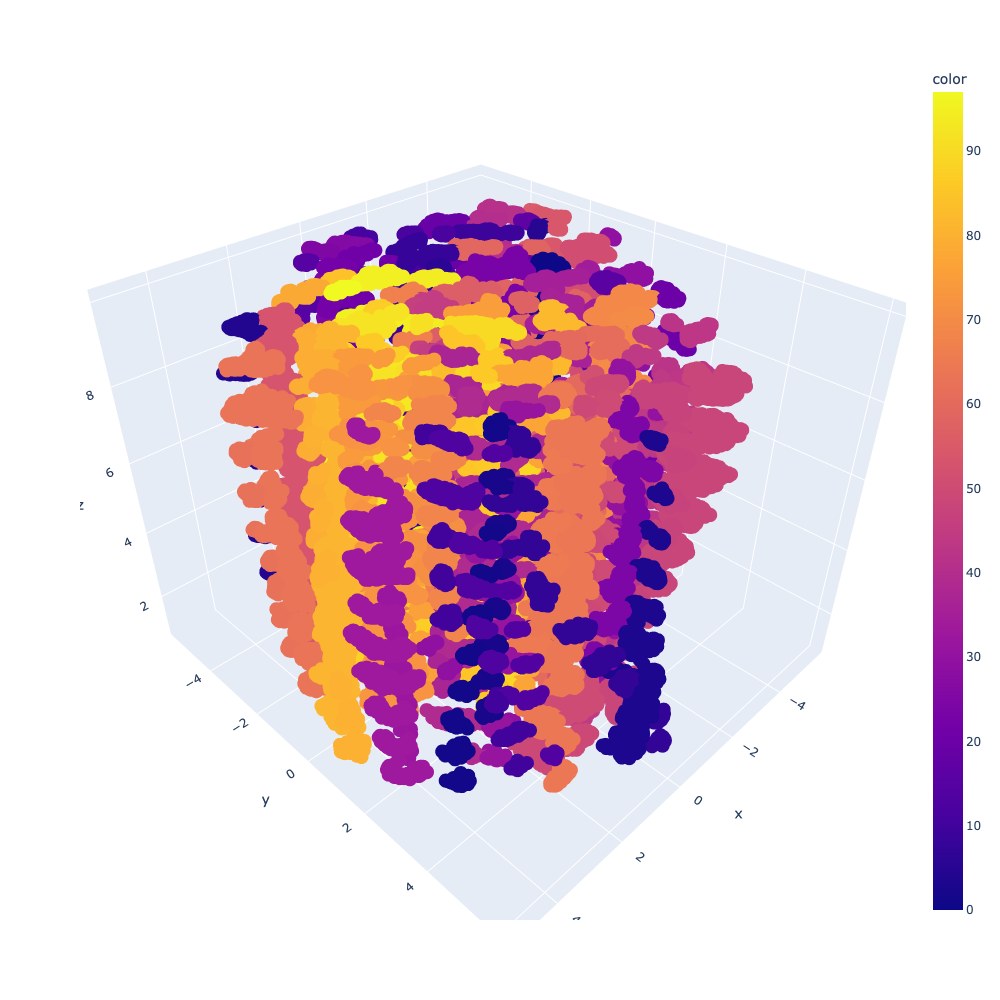}
    \caption{BERT | DBLP}
  \end{subfigure}
  \caption{Evolving Clusters: \textnormal{These clusters (that share the same color) are the results of the sequential clustering of documents in each time frame, which are aligned with their consecutive time frames. The representation of these Evolving Clusters can help explore the evolution of topics over time as described in \Cref{fig:musk}, \Cref{fig:wordsdblp}, and \Cref{fig:wordsnyt}.}}
    \label{fig:3d}
\end{figure*}

To obtain a representation of the temporal evolution of these clusters, we apply a final alignment step that generates document clusters based on their embedding similarity. The result is a set of \textit{evolving document clusters} that contain documents that are semantically similar to documents in the same time period, but also to other documents in previous or subsequent time periods:
{\definition Let $\{D_i^t\}$ be the set of local document clusters obtained by $\PC$.
Cluster Alignment $\CA: D \mapsto \{ D_k \}$ generates $m$ subsets of documents $D_k\subseteq D$ for $k=1$ to $m$ and where $D_k$ is the \textit{union} of a set of local document clusters (topics) $D_i^t$ with \emph{similar embedding clusters $Y_i^t$}.
}
We call $D_k$ an evolving cluster (EC) and $D_i^t$ a local document cluster of $D_k$. We can use different cluster linkage measures (single, average, centroid, complete) to estimate the similarity of the embedding clusters $Y_i^t$ (also note that since the embeddings of the local document clusters of $D_k$ are similar, all the documents in $D_k$ are similar as well). \Cref{fig:3d} shows two examples of evolving clusters in two datasets with two different embedding methods.   

\subsubsection{Implementation}
This layer, similar to the other two layers of ANTM, has high flexibility and scaleability to be implemented using several techniques and algorithms. As shown in \Cref{fig:arc}, we implemented the proposed overlapping sliding algorithm with density-based clustering techniques~\cite{ghosal2020short}. 
Density-based clustering algorithms identify clusters in a data set based on the density of points in feature space. These algorithms are particularly useful for identifying clusters of varying densities and shapes. They are also sensitive to the parameters used to compute the density and to the presence of noise or outliers in the data. In addition, these algorithms are useful when the clusters have arbitrary shapes and when the number of clusters is not known in advance. This feature provides an advantage over non-parametric PDTMs that require a fixed number of topics as input. The Aligned Clustering is developed in the following three steps.

\paragraph{Aligned Dimension Reduction}
In order to overcome the limitations posed by the curse of dimensionality in density-based clustering algorithms, it is necessary to reduce the dimensionality of document embeddings. The solution proposed is to utilize AlignedUMAP~\cite{islam2022manifold}, a method that aligns sequences of UMAP~\cite{mcinnes2018umap} embeddings based on the relationships between the data frames. This allows for local dimensionality reduction of document vectors within each time frame, taking into account their semantic patterns, thereby ensuring that they remain within the same vector space and maintain continuity.


\paragraph{Sequential Local Clustering}
%

%
Afterward, a density-based clustering algorithm is performed on each data frame to group them sequentially into a set of clusters. An effective approach is HBDSCAN~\cite{campello2013density}, which extends DBSCAN~\cite{ester1996density} by converting the model into a hierarchical clustering algorithm and then using a technique to extract a flat clustering based on the stability of the clusters. As shown in \Cref{fig:clusterD2V,fig:bertDBLP}, these clusters describe a concept within a time frame.

\paragraph{Cluster Alignment}
Finally, this step aligns consecutive clusters achieved from the previous step. Since the document embedding vectors of all time frames are in the same space, we can aggregate each cluster by considering their centroids. We then use the HDBSCAN clustering technique as linkage method~\cite{jarman2020hierarchical} to align document clusters from different periods by identifying the clusters of centroids. An example of the output for this step is shown in \Cref{fig:3d}.

\subsection{Representation Layer}
The representation layer is responsible for providing word representations for each cluster in each time frame within documents. The goal of this layer is to identify the most relevant terms or phrases that characterize a particular concept within a collection of documents. These terms or phrases can be used to summarize the main ideas or themes of the document clusters. 
{\definition The topic Representation Layer ($\RL$) computes a list of $m$ terms $\{ t_{ij}^r \}_{r=1}^m$ that describe the semantic contents of each document cluster $D_{ij}$.


There exist several ways to represent a set of documents within a set of extracted words. One way is to use the joint word and document embedding over a document cluster and find the $m$ closest words from the centroid of the document embeddings. Another way is to use a method called class-based Term Frequency-Inverse Document Frequency (c-TF-IDF)~\cite{grootendorst2022bertopic}, which is a variation of TF-IDF that takes into account the class labels of documents. c-TF-IDF weights the terms in a document not only by their frequency but also by their relevance to a particular class. c-TF-IDF ensures that each word of a topic representation occurs in at least one document of that topic. Examples of outputs for this step are shown in \Cref{fig:wordsdblp,fig:wordsnyt}. 

\begin{figure*}[t]
\centering
\includegraphics[width=0.8\textwidth]{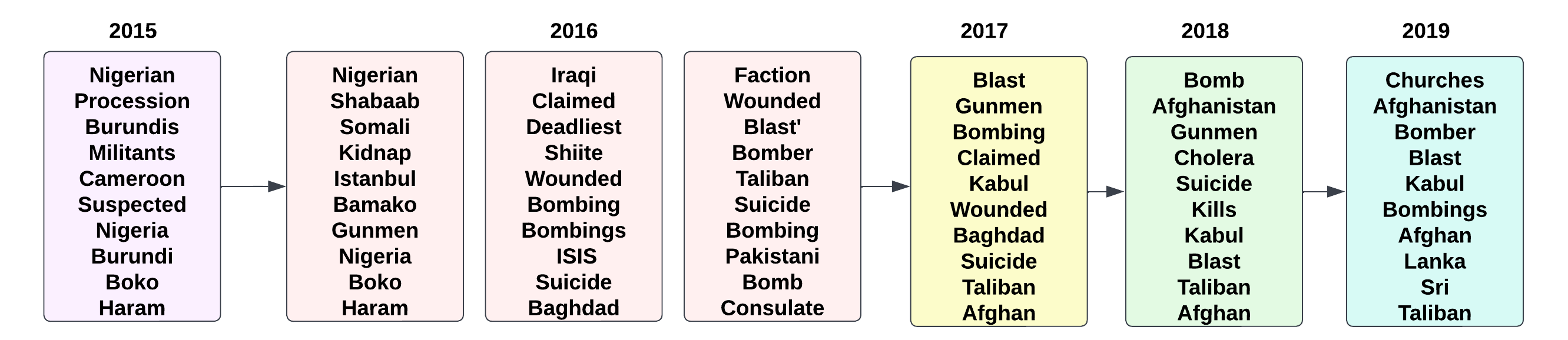}
\caption{Evolution of New York Times News on Foreign Terrorist Activities. \textnormal{The interesting transition of news from Boko, Harram, Nigeria in 2015 to ISIS, Iraq, Baghdad in 2016 and later Kabul, Taliban, Afghanistan in 2017 to 2019.}}
\label{fig:wordsnyt}
\end{figure*}

\begin{figure*}[hbtp]
\centering
\includegraphics[width=\textwidth]{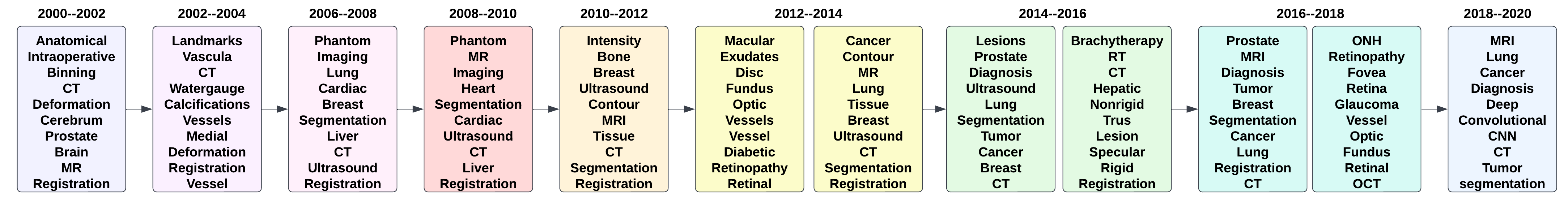}
\caption{Evolution of Computer Science Research on Medical Science based on DBLP documents. \textnormal{The words CNN, Lung, and Diagnosis appear in the period 2018 to 2020, which coincides with the pandemic and advances in computer vision research.} }
\label{fig:wordsdblp}
\end{figure*}




\section{Experiments}
\label{sec:experiment}
The aim of this experiment is to implement ANTM and discover its optimal hyperparameters in terms of the quality of the generated topics. We will then compare ANTM with baselines (DTM\cite{blei2006dynamic}, DETM\cite{dieng2019dynamic} and BERTopic\cite{grootendorst2022bertopic}) and illustrate its capability for exploratory evolutionary analysis.

\subsection{Datasets}
Four datasets are used in our experiments for different purposes. The first dataset is the DBLP~\cite{ley2002dblp} archive of 168K scientific articles (title and abstract) published between 2000 and 2020. 
DBLP is used to compare the proposed method with DTM~\cite{blei2006dynamic} and BERTopic~\cite{grootendorst2022bertopic}. The second dataset is a 16K sample of documents (title and abstract) extracted from arXiv~\cite{clement2019use} and published between 2006 and 2022. This dataset is used to compare ANTM with two settings (Data2Vec and BERT) against DETM~\cite{dieng2019dynamic} and BERTopic. 
The third dataset is a collection of 19K tweets~\cite{shahzad2022price} published by Elon Musk between 2017 and the end of 2022. 
The last dataset is a collection of 210K articles~\cite{pinter2020nytwit} from the New York Times published between 2012 and 2022. 
Elon Musk's tweets and the New York Times dataset are used for exploratory evolutionary analysis of evolving topics based on the proposed method. The statistics of the datasets are summarized in \Cref{tb1}.

\begin{table}[tbh]
\small
\caption{The statistics of the datasets used in the experiment}
\label{tb1}
\centering
 \begin{tabular}{||c|| c| c| c| c||} 
 \hline
 Dataset & Documents & Tokens & Vocabulary & Date Range\\ 
 \hline\hline
 DBLP & 168K & 18.5M & 144K & 2000-2020 \\ 
\hline
arXiv & 16K & 1.17M & 12.6K & 2006-2022  \\ 
 \hline
EM's Tweets & 19K & 91.5K & 4K & 2017-2022 \\
 \hline
NYT News & 210K & 5M & 61K & 2012-2022  \\ 
 \hline
 \end{tabular}
\end{table}

\subsection{Baselines}
The performance of ANTM is compared with three main dynamic topic models. The first model is DTM\cite{blei2006dynamic}, which has been widely used for various applications such as public opinion tracking\cite{yan2020exploring}. The second model is DETM~\cite{dieng2019dynamic}, which extends DTM with word embeddings to improve topic quality and performance. The last model is BERTopic\cite{grootendorst2022bertopic}, which is an LLM-based topic model that uses a static process for clustering document embeddings while dynamically representing topics. 

\subsection{Evaluation Metrics}
Since the number of topics generated by a dynamic topic model is not a sufficient indicator of quality or superiority over other models, we evaluate the performance of the topic models in terms of human interpretability and diversity using Topic Coherence (TC)~\cite{newman2010automatic,roder2015exploring,rahimi2023contextualized} and Topic Diversity (TD)~\cite{hashimoto2021analyzing,dieng2019dynamic}.

\paragraph{Topic Coherence (TC)} 
 This metric indicates the interpretability of a topic containing $m$ words and by estimating that a topic is highly coherent if it is represented by words that tend to occur in the same documents. TC, as defined in \cite{roder2015exploring} with co-occurrence value $C_V$, is a variation of Normalized Point-wise Mutual Information (NPMI)\cite{bouma2009normalized} that average co-occurrence value between each pair of words ($t_i^r$ and $t_i^s$) in topic $i$ over $N$ topics.



\paragraph{Topic Diversity (TD)} 
Topic diversity estimates the diversity of the topic representations within a given set of topics. For this metric, we use the Proportion of Unique Words (PUW) method as suggested by~\cite{dieng2020topic} defined as the vocabulary size divided by the total number of words within a set of topics. Topics within a low-diversity topic set share many words, whereas a high-diversity topic set contains topics that have few words in common.

%

%



\subsection{Experimental Setup}

\paragraph{ANTM}
Two embedding configurations are used in the contextual embedding layer ($\CEL$) of ANTM: Data2Vec (ANTM-Data2Vec) and BERT (ANTM-BERT). For evaluating ANTM-Data2Vec, we ran pre-trained Data2Vec (\textit{facebook/data2vec-text-base}\cite{huggingface}) on the document abstracts of the arXiv dataset. Similarly, we ran BERT (\textit{all-mpnet-base-v2}~\cite{reimers2019sentence}) on the titles and abstracts of DBLP, the textual content of Elon Musk's tweets, and the descriptions of NYT articles to obtain document embedding regarding the semantic context of the datasets for evaluating ANTM-BERT. 
In both configurations, each document is mapped to a 768-dimensional dense vector space. We then split the embedded vectors into a series of time frames to explore the dynamic change within the content of documents. The setting information of the segmentation step is summarized in \Cref{tb2}. These segmentation values, as suggested in \cite{anderson2012towards}, provide a comprehensive view of the data and ensure that changes over time are captured in the analysis.

\begin{table}[tbh]
\small
\caption{The summary of segmentation setting}
\label{tb2}
\centering
 \begin{tabular}{||c|| c| c| c||} 
 \hline
 Dataset \& Time Frame Sizes & Length & Overlap & \#Frames \\ 
 \hline\hline
 DBLP Documents & 3 Years & 1 Year & 10  \\ 
 \hline
 arXiv Documents& 3 Years & 1 Year & 8  \\ 
 \hline
New York Times Articles  & 3 Years & 1 Year & 10   \\ 
\hline
Elon Musk's Tweets & 6 Months & 2 Months & 16  \\
\hline
\end{tabular}
\end{table}

The next step is to find optimal hyperparameters for dimension reduction (using AlignedUMAP) and sequential document clustering (using $\PC$).

In this regard, we conduct a grid search study on various parameters of ANTM in order to assess the individual contributions of each parameter toward the overall performance of the DBLP dataset. In this study, we carefully selected the following ranges to explore different aspects: UMAP dimensions were set to [2, 3, 4, 5], UMAP Neighbor numbers ranged from [10, 15, 20, 50, 100], and minimum cluster sizes in $\PC$ varied across [10, 15, 20, 50, 100]. This comprehensive selection allows us to thoroughly examine the impact of these parameters. For each combination, we score the results based on Topic Quality (TC$\times$TD) and the number of clusters they produce since Topic quality by itself is not sufficient to find the optimal configuration, because some rare cases of configuration contain only one large topic with high diversity and coherence. To avoid such a case, we define the topic $score$ by normalizing the topic quality relative to the number of topics as follows.

\begin{equation}
    \text{Score}=\frac{1}{n}\sum_{i=0}^{n-1}\text{TC}_i \times \text{TD}_i\times \frac{T_i}{\text{T}_i^{max}}
\end{equation}
where $\text{Score}$, $\text{TC}_i$, and $\text{TD}_i$ represent the quality score, topic coherence, and topic diversity in time frame $i$. Furthermore, let $T_i$ denote the number of topics in time frame $i$, and $\text{T}_i^{max}$ represent the maximum number of topics achieved through experimentation across all combinations of settings.

\Cref{fig:grid}(a) and \Cref{fig:grid}(b) display box plot representations, illustrating the distribution of the number of clusters ($T_i$) and topic quality ($Q_i$) per time frame on a sample of DBLP dataset (4 slices totally, 5000 documents per year ranging from 2005 to 2020), respectively. 
\begin{figure*}[ht]
  \centering
  \begin{subfigure}[b]{0.4\textwidth}
   \centering
    \includegraphics[width=\textwidth]{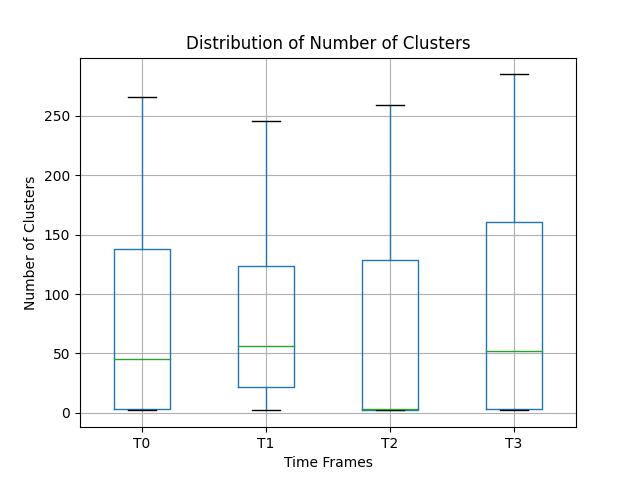}
    \caption{Number of Clusters per Period}
  \end{subfigure}
   \hfill
  \begin{subfigure}[b]{0.4\textwidth}
   \centering
    \includegraphics[width=\textwidth]{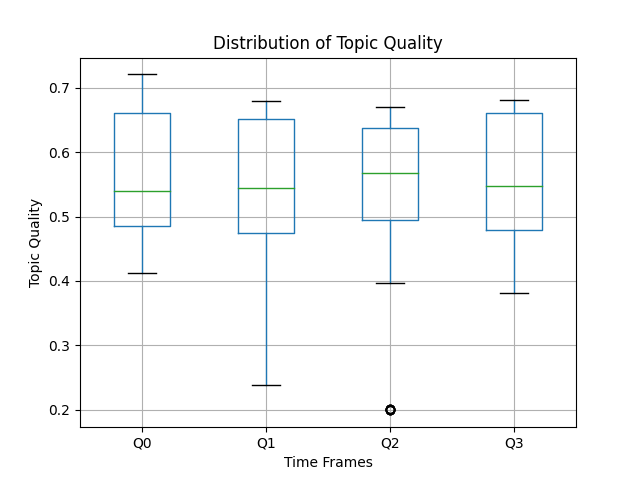}
    \caption{Topic Quality per Period}
  \end{subfigure}
  \caption{Box Plot of distributions of the grid search study on a sample of DBLP.}
    \label{fig:grid}
\end{figure*}
Lastly, we compute the average quality scores across all time frames and arrange them in ascending order based on this aggregated value. \Cref{tab:abl} showcases the top-5 scoring settings obtained from this study conducted on the DBLP dataset, shedding light on the significance of these parameters. 
\begin{table*}[tbh]
\caption{The grid search study about how each parameter contribute to the overall performance of the experiment}
\centering
\small
\begin{tabular}{|ccc|cccc|cccc||c|}
\toprule
   UMAP \#D &  UMAP \#N &  PC Size &  $T_0$ &  $T_1$ &  $T_2$ &  $T_3$ &  $Q_0$ &      $Q_1$ &      $Q_2$ &      $Q_3$ &    Score  \\
\midrule
             2,3,4,5 &                10 &                           10 & 266 & 246 & 259 & 285 & 0.41 & 0.41 & 0.43 & 0.42 & 0.42 \\
             2,3,4,5 &                15 &                           10 & 231 & 234 & 229 & 234 & 0.42 & 0.43 & 0.43 & 0.42 & 0.37 \\
             2,3,4,5 &                50 &                           10 & 176 & 162 & 188 & 200 & 0.44 & 0.45 & 0.45 & 0.46 & 0.31 \\
             2,3,4,5 &                15 &                           15 & 169 & 151 & 147 & 161 & 0.48 & 0.50 & 0.50 & 0.50 & 0.29 \\
             2,3,4,5 &                10 &                           20 & 131 & 124 & 124 & 165 & 0.52 & 0.52 & 0.54 & 0.50 & 0.26 \\
        
\bottomrule
\end{tabular}
\label{tab:abl}
\end{table*}
The results reveal that varying the UMAP dimensions (2, 3, 4, 5) did not have any noticeable effect on the outcomes. However, utilizing a UMAP size of 10 and a minimum cluster size of 10 yielded the highest scores.

Accordingly, we chose our parameters as follows. The cosine similarity metric, with 5-dimensional output, was chosen for the dimensionality reduction setting of AlignedUMAP. We then performed HDBSCAN on each time frame with Euclidean distance and a minimum size of $10$ documents per cluster to create a set of semantically similar document clusters for each period (\Cref{fig:clusterD2V,fig:bertDBLP}). Since all document embeddings are in the same vector space, we could then align the generated clusters by again using HDBSCAN on the centroid of all document clusters with Euclidean distance and a minimum number of $2$ documents to obtain evolving clusters for each dataset (\Cref{fig:3d}). Finally, we represented the documents of each cluster with a set of $m=10$ words using the c-TF-IDF (\Cref{fig:musk,fig:wordsdblp,fig:wordsnyt}).

\paragraph{Baselines}
 DTM is configured with 50 and 100 topics on the titles and abstracts of the DBLP dataset. In addition, DETM is configured with 100 topics and 50 epochs on the abstract of the arXiv dataset. The number of topics for DTM and DETM is determined based on the number of topics generated by ANTM in an unsupervised manner. Furthermore, we ran BERTopic with the same sentence transformer model used for the proposed model on both DBLP and arXiv datasets. For the rest of the hyperparameters of BERTopic, we used the default configurations. Finally, the number of word representations for each topic model is considered equal to 20 for arXiv documents and 10 for the rest of the datasets.

\section{Results}
\label{sec:comparison}
The proposed model based on BERT document embedding generated 100 evolving topics on the DBLP dataset. For the arXiv dataset, the proposed model based on Data2Vec and BERT generated 130 and 110 evolving topics, respectively. On the other hand, BERTopic generated 500 global topics on the DBLP dataset and 120 global topics on the arXiv dataset. 

Using TC and TD metrics, we can analyze the results obtained by the proposed model and evaluate the models from two perspectives. First, we can look periodically at each time frame and observe the temporal quality of all topics in terms of coherence and diversity of word representations. This analysis compares the temporal ability of dynamic topic models to describe temporal topics in a diverse and coherent manner. The second perspective is to analyze the quality of the word representations presented over each topic. This analysis compares the ability of dynamic topic models to explore the evolution of each dynamic topic over time.




\subsection{Period-Wise Quality Analysis}
The goal of the period-wise analysis is to determine whether aligned document clustering, applied separately for each time frame, results in better topic quality than the competing BERTopic, which uses a global document clustering procedure. As shown in \Cref{fig:pwq:a,fig:pwq:b,fig:pwq:c}, the first period-wise evaluation is based on the DBLP dataset, which has the largest number of documents and vocabulary size among our datasets. \Cref{fig:pwq:a,fig:pwq:b,fig:pwq:c,fig:pwq:d} compare ANTM-Bert with the other models applied to DBLP. On \Cref{fig:pwq:a}, ANTM-BERT has a higher topic coherence ranging from 66\% to 70\% compared to DTM and BERTopic which remain below 63\%. On the other hand, on  \Cref{fig:pwq:b}, the topic diversity of ANTM-BERT is less than 5 percent lower than BERTopic and about 10 to 15 percent higher than DTM. Overall, the quality of ANTM-BERT, which is the product of TC and TD, is still higher than DTM and BERTopic based on the DBLP dataset. This evaluation can illustrate the capability of two ADTMs over DTM, which is a non-embedding based probabilistic dynamic topic model. Furthermore, it validates our argument that local clustering using the Aligned Clustering method, preserves more information and is more descriptive in explaining temporal topics. 

As shown in \Cref{fig:pwq:d,fig:pwq:e,fig:pwq:f},  ANTM-Data2Vec and ANTM-Bert is compared against embedding-based models on the arXiv dataset. On \Cref{fig:pwq:e}, both configurations of ANTM produce topics that are 25\% more diverse than those produced by DETM and about 10\% less diverse than those obtained by BERTopic. On the other hand, on \Cref{fig:pwq:d}, the topics of ANTM-Data2Vec are about 20\% more coherent than the topics of ANTM-BERT and BERTopic and about 15\% less coherent than the topics produced by DETM. Since the quality of topic models is defined based on an aggregation of topic coherence and topic diversity, i.e., diversity is valuable when there is coherence and vice versa, ANTM-Data2Vec has the highest quality score compared to BERTopic and DETM (\Cref{fig:pwq:f}). This evaluation also shows that ANTM-Data2Vec is more effective than ANTM-BERT on arXiv. Moreover, it reinforces the argument that the Aligned Clustering method preserves more information and is more descriptive in explaining temporal topics than ADTMs with global clustering. 

\begin{figure*}[htbp]
  \centering
  \begin{subfigure}[b]{0.33\textwidth}
    \includegraphics[width=\textwidth]{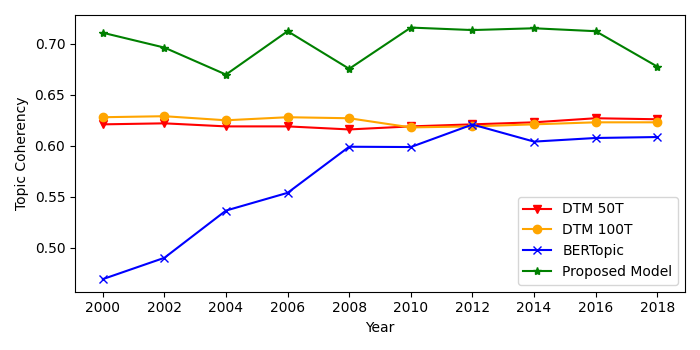}
    \caption{Topic Coherence Metric on DBLP dataset}
    \label{fig:pwq:a}
  \end{subfigure}
  \hfill
    \begin{subfigure}[b]{0.33\textwidth}
    \includegraphics[width=\textwidth]{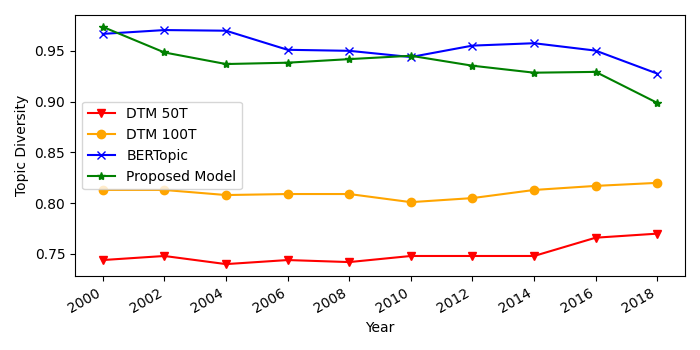}
    \caption{Topic Diversity Metric on DBLP dataset}
    \label{fig:pwq:b}
  \end{subfigure}
  \hfill
  \begin{subfigure}[b]{0.33\textwidth}
    \includegraphics[width=\textwidth]{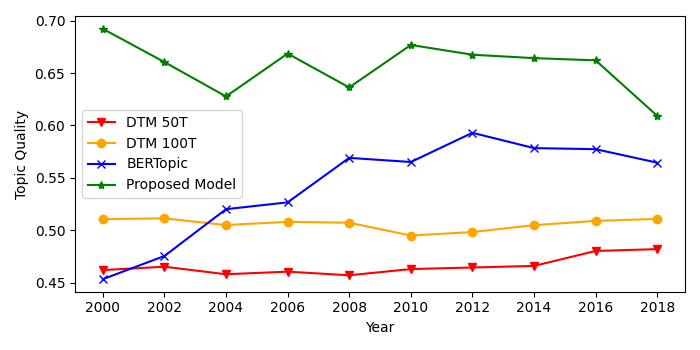}
    \caption{Topic Quality Metric on DBLP dataset}
    \label{fig:pwq:c}
  \end{subfigure}

  \hfill
\begin{subfigure}[b]{0.33\textwidth}
    \includegraphics[width=\textwidth]{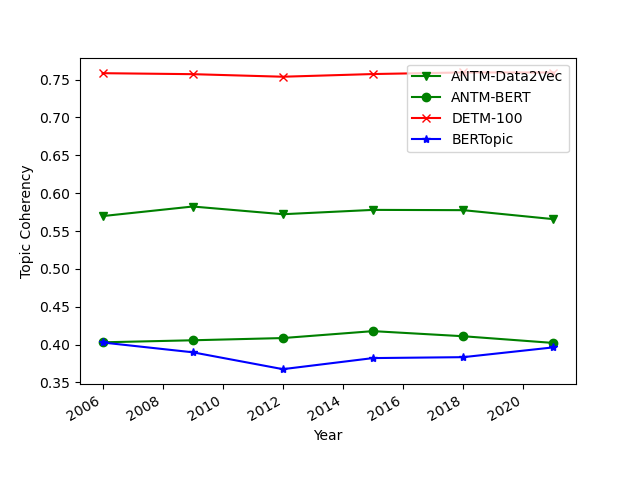}
    \caption{Topic Coherencc Metric on arXiv dataset}
    \label{fig:pwq:d}
  \end{subfigure}
  \hfill
    \begin{subfigure}[b]{0.33\textwidth}
    \includegraphics[width=\textwidth]{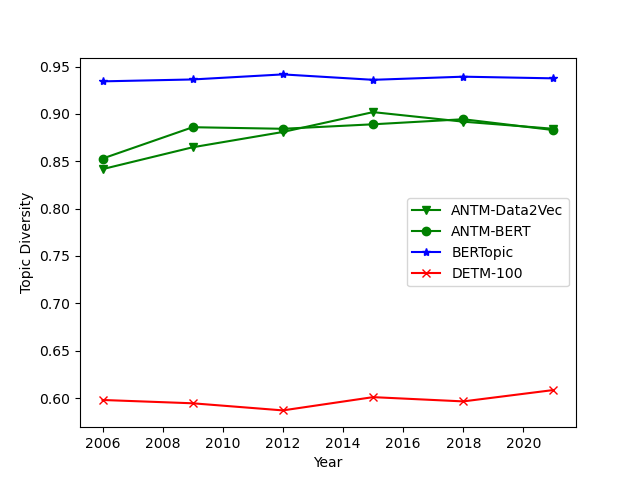}
    \caption{Topic Diversity Metric on arXiv dataset}
    \label{fig:pwq:e}
  \end{subfigure}
  \hfill
  \begin{subfigure}[b]{0.33\textwidth}
    \includegraphics[width=\textwidth]{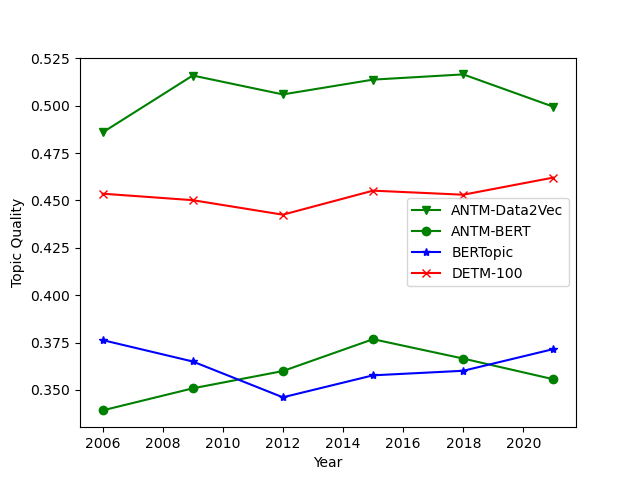}
    \caption{Topic Quality Metric on arXiv dataset}
    \label{fig:pwq:f}
  \end{subfigure}

\caption{Period-wise comparison between Dynamic Topic Models}
\label{fig:pwq}
\end{figure*}

\subsection{Topic-Wise Quality Analysis}
The objective of topic-wise analysis is to assess whether an evolving topic, which contains topics that are semantically close to each other, can nevertheless be diversified. In each model, topic coherence and topic diversity are calculated separately for each evolving topic. As shown in \Cref{fig:twq}, we have sorted these values and plotted their distributions across each model on the DBLP and arXiv datasets. These figures contain multiple observations. 
\begin{figure*}[htb]
  \centering
  \begin{subfigure}[b]{0.22\textwidth}
    \includegraphics[width=\textwidth]{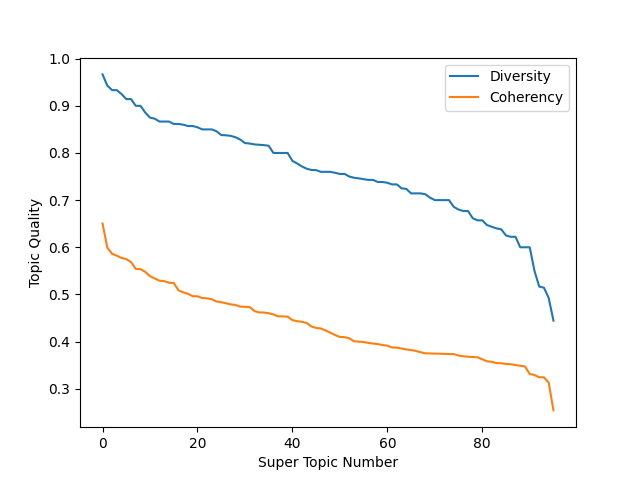}
    \caption{ANTM-BERT on DBLP}
    \label{fig:twq:a}
  \end{subfigure}
  \hfill
    \begin{subfigure}[b]{0.22\textwidth}
    \includegraphics[width=\textwidth]{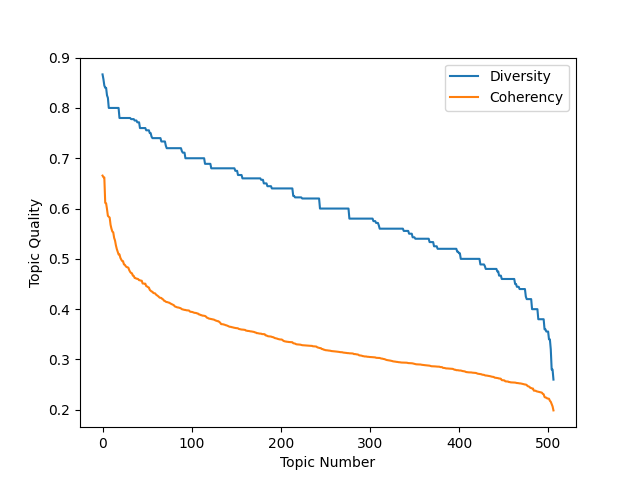}
    \caption{BERTopic on DBLP}
    \label{fig:twq:b}
  \end{subfigure}
  \hfill
      \begin{subfigure}[b]{0.22\textwidth}
    \includegraphics[width=\textwidth]{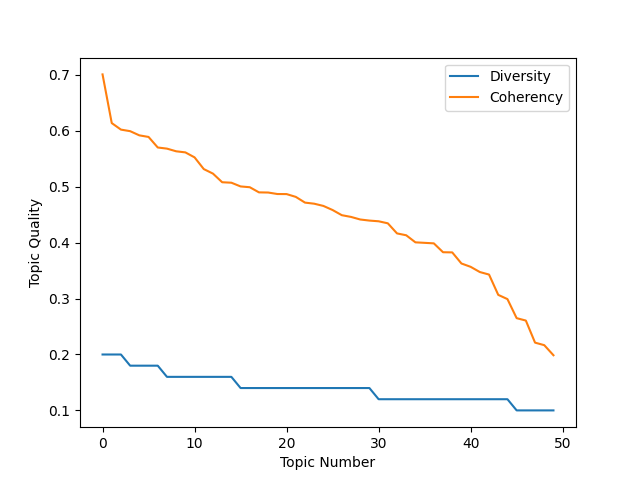}
    \caption{DTM with 50T on DBLP}
    \label{fig:twq:c}
  \end{subfigure}
  \hfill
  \begin{subfigure}[b]{0.22\textwidth}
    \includegraphics[width=\textwidth]{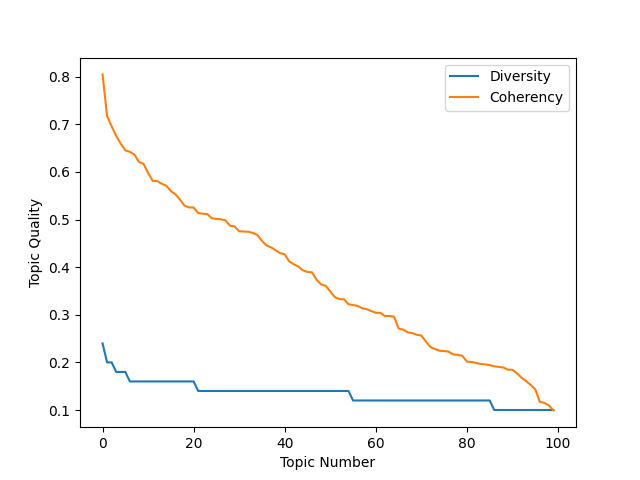}
    \caption{DTM with 100T on DBLP}
    \label{fig:twq:d}
  \end{subfigure}
  \hfill
    \begin{subfigure}[b]{0.22\textwidth}
    \includegraphics[width=\textwidth]{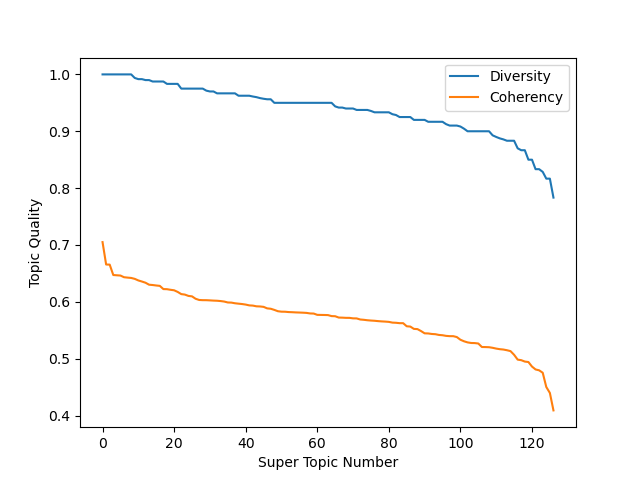}
    \caption{ANTM-Data2Vec on arXiv}
    \label{fig:twq:e}
  \end{subfigure}
  \hfill
    \begin{subfigure}[b]{0.22\textwidth}
    \includegraphics[width=\textwidth]{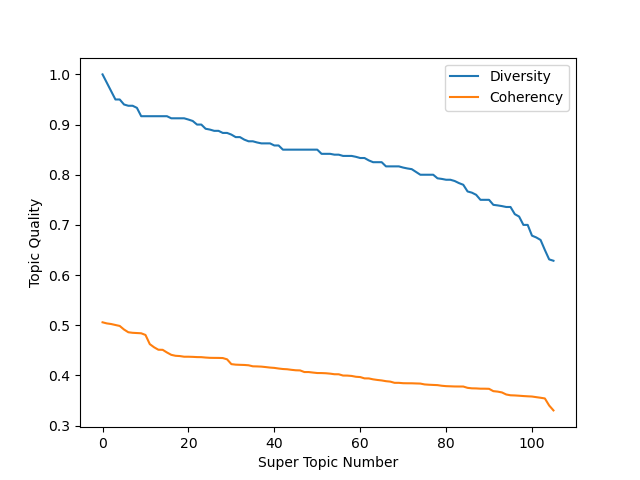}
    \caption{ANTM-BERT on arXiv}
    \label{fig:twq:f}
  \end{subfigure}
  \hfill
      \begin{subfigure}[b]{0.22\textwidth}
    \includegraphics[width=\textwidth]{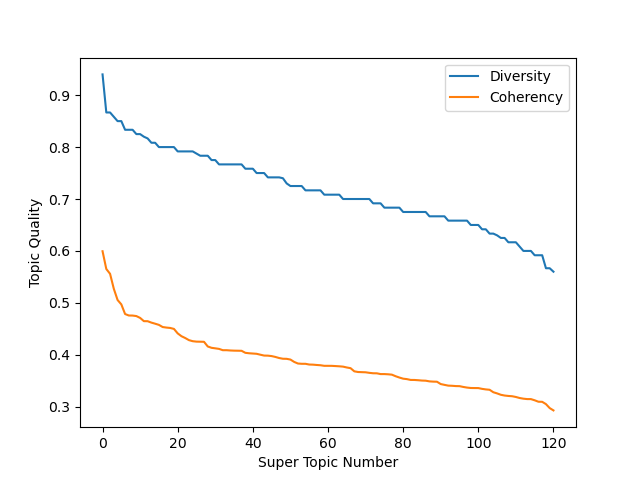}
    \caption{BERTopic on arXiv}
    \label{fig:twq:g}
  \end{subfigure}
  \hfill
  \begin{subfigure}[b]{0.22\textwidth}
    \includegraphics[width=\textwidth]{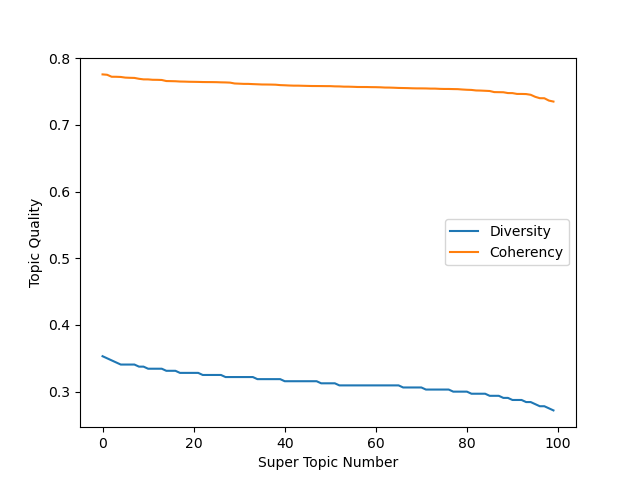}
    \caption{DETM with 100T on arXiv}
    \label{fig:twq:h}
  \end{subfigure}
 
  \caption{Topic-wise distribution of quality metrics over Dynamic Topic Models }
   \label{fig:twq}
\end{figure*}

First, as explained in \Cref{sec:related}, the use of document and word embeddings increases the diversity of evolving topics (\Cref{fig:twq:a,fig:twq:b,fig:twq:c,fig:twq:e,fig:twq:f,fig:twq:g}). The high diversity of embedding-based models can be explained by the fact that evolving topics may regroup documents with semantically similar but syntactically different words (e.g., bus, car, bike). Since topic diversity is based on the syntactic (c-TF-IDF) topic representations, this leads to more heterogeneous local topic representations within evolving topics and might explain why the evolving topics obtained by BERT and Data2Vec have a higher diversity than those obtained by DTM (\Cref{fig:twq:c,fig:twq:d}) and DETM (\Cref{fig:twq:h}). 

On the other hand, the average topic coherence of topics in evolving topics is more difficult to explain. First, we can observe that DETM (\Cref{fig:twq:h}) achieves a very high and almost constant topic coherence compared to the other methods. This seems to be the result of the use of word embeddings during the Bayesian model generation process.  Among the approaches based on document embeddings, ANTM-Data2Vec achieves better average topic coherence (and topic diversity) (\Cref{fig:twq:e}) than the other approaches. 


\section{Conclusion}
\label{sec:conclusion}
Existing dynamic topic models ignore certain temporal variations of evolving topics by configuring a global structure for dynamic topics, such as the same number of document clusters in each period. Moreover, some of these models are not scalable to large archives with large dictionaries and are not capable of handling short texts for applications such as social network analysis. These limitations directly affect the observation of topic evolution and reduce the interpretability and diversity of evolving topics, which are sequentially represented. In this paper, we proposed an algorithmic family of dynamic neural topic models called Aligned Neural Topic Models (ANTM), which combines novel data mining algorithms to provide a modular framework for discovering evolving topics. ANTM outperforms the state-of-the-art dynamic topic models in terms of interpretability and diversity of topic words based on a series of experiments on four distinct datasets.
As future work, we plan to define a metric for assessing the quality of topic alignments and their ability to detect topic emergence or fading. We will also investigate embedding-based methods for inferring more comprehensive topic labels from document clusters.

\section*{Acknowledgement}
We gratefully acknowledge the Sorbonne Center for Artificial Intelligence (SCAI) for partially funding this research through a doctoral fellowship grant. We'd like to thank our students, Lucie Chen, Mohamed Allaa Eddine Boutaleb, and Mouloud Samy Nehlil for their assistance. This paper is dedicated to the brave women of Iran.


\bibliographystyle{ieeetr}
\bibliography{references}

\end{document}